\documentclass[journal,twoside]{IEEEtran}

\usepackage{amsmath}
\usepackage{amsfonts}
\usepackage{amssymb}
\usepackage{hyperref}
\usepackage{subfig}
\usepackage{booktabs}
\usepackage{cite}
\usepackage{graphicx}

\ifCLASSINFOpdf
\else
\fi
\begin{document}
%
\title{MODEL: Motif-based Deep Feature Learning for Link Prediction}
%
%
%

\author{Lei Wang, Jing Ren, Bo Xu, Jianxin Li, Wei Luo, and Feng~Xia,~\IEEEmembership{Senior~Member,~IEEE}\\
	\thanks{Manuscript received August 31, 2019; revised November 18, 2019; accepted December 21, 2019. This work is partially supported by National Natural Science Foundation of China under Grant No. 61872054 and the Fundamental Research Funds for the Central Universities (DUT19LAB23). This research also is partially supported by Australian Research Council (ARC) Linkage Project with Grant Number- ARC LP180100750. \emph{(Corresponding author: Bo Xu; e-mail: boxu@dlut.edu.cn)}}
	\thanks{L. Wang, J. Ren, and B. Xu are with Key Laboratory for Ubiquitous Network and Service Software of Liaoning Province, School of Software, Dalian University of Technology, Dalian 116620, China.}
	\thanks{J. Li and W. Luo are with School of Information Technology, Deakin University, Burwood, Melbourne, VIC 3125, Australia.}
	\thanks{F. Xia is with School of Science, Engineering and Information Technology, Federation University Australia, Ballarat, VIC 3353, Australia.}}

%
%

\markboth{IEEE Transactions on Computational Social Systems,~Vol.~0, No.~0, August~2019}%
{Wang \MakeLowercase{\textit{et al.}}: MODEL: Motif-based Deep Feature Learning for Link Prediction}
%



\maketitle

\begin{abstract}
Link prediction plays an important role in network analysis and applications.
Recently, approaches for link prediction have evolved from traditional similarity-based algorithms
into embedding-based algorithms.
However, most existing approaches fail to exploit
the fact that real-world networks are different from random networks.
In particular, real-world networks are known to contain motifs, natural network building blocks 
reflecting the underlying network-generating processes. 
In this paper, we propose a novel embedding algorithm that incorporates network motifs to capture higher-order
structures in the network.
To evaluate its effectiveness for link prediction,
experiments were conducted on three types of networks:
social networks, biological networks, and academic networks.
The results demonstrate that our algorithm outperforms both the traditional similarity-based algorithms (by 20\%)
and the state-of-the-art embedding-based algorithms (by 19\%).
\end{abstract}


%
\IEEEpeerreviewmaketitle

\section{Introduction}
	\IEEEPARstart{C}{omplex} networks occur in many natural and social settings, examples including academic networks \cite{xia2017big}, biological networks \cite{38}, social networks \cite{40,41}, and vehicular networks \cite{37}.
A network contains both vertices representing entities and links representing relationships between entities.
It is a well-known problem that many networks generated from relational databases
suffer from missing links.
\emph{Link prediction} aims to identify such missing links in a network \cite{1,4,6,18}.
It is an active research area that has many useful applications:
Through link prediction, we can recommend a scientific paper for scholars in academic networks \cite{xia2017big}, build social recommender systems \cite{43, 45}, and  discover currently unknown protein-protein interactions in biological networks \cite{44}.

In link prediction, a score is assigned to each pair of unconnected vertices based on their features.
A higher score implies a higher probability of a missing link connecting the two vertices.
Traditional similarity-based algorithms for link prediction rely on hand-crafted features,
such as Common Neighbors and Jaccard's Coefficient \cite{1}.
However, when creating hand-crafted features,
we need to select informative features and consider specific domain knowledge.
To solve this problem, the idea of automatically learning features,
known as network embedding \cite{2, cui2018survey}, has emerged.
The main idea of network embedding is to map each vertex in a network to a vector
in a \textit{d}-dimensional space,
with the correlation of vectors reflecting the correlation of the corresponding vertices.
For example, the cosine of the vectors of two vertices can indicate the strength of their similarity or
correlation.

Although existing algorithms for link prediction achieve good performance,
they face the following challenges:\par
\textbf{(1) Higher-order structures:} Real-world networks usually exhibit higher-order structures.
This is, the networks contain a large number of connected subgraphs that occur more frequently
than in random networks.
The well-known Science paper \cite{9} has demonstrated that such subgraphs, known as network motifs, are
basic network "building blocks" reflecting specific types of networks.
A lot of studies in network analysis \cite{10,17} have confirmed that leveraging subgraphs
can improve experimental performance.
In particular, frequent subgraphs have been successfully applied to improve
whole-graph embedding \cite{doi:10.1137/1.9781611975321.35}.
However, to our knowledge, network motifs have not been used to improve link prediction.

\textbf{(2) Weak ties:} The number of common neighbors between two vertices is informative of measuring their similarity. However, some vertices exhibit a weak tie, which means they share a few or no common neighbors \cite{onnela2007structure}. We can find this phenomenon where two vertices reside in two different communities \cite{47}. If we only concentrate on common neighbors, it will be difficult to discover the potential edge between two vertices that exhibit a weak tie. \par
\textbf{(3) Complex network structures:} Many networks have complex and highly non-linear structures,
as demonstrated in \cite{46}.
Existing embedding-based algorithms rely on linear and shallow models, such as matrix factorization,
which are unable to capture the complex structure of networks \cite{6}.\par

To sum up, the majority of existing algorithms fail to capture higher-order structures and potential weak ties since they rely on only pair-wise similarity and immediate neighbors. 
This problem can be addressed using special subgraphs (such as motifs) that preserve higher-order connectivity patterns \cite{9}. 
Such higher-order connectivity patterns are often crucial for link prediction.  
In addition, existing algorithms, for example HONE \cite{18},  often learn features leveraging only shallow models. 
Often a complex model is necessary to capture complex network structures to achieve optimal link prediction \cite{46}. 

In this paper, we propose a \textbf{MO}tif-based \textbf{DE}ep feature learning algorithm for \textbf{L}ink prediction named MODEL. 
The proposed algorithm uses motifs to automatically learn vectorized features. When learning vector representations, we hope that learned vectors of these vertices composing a motif close to each other, and ensure that overlapping motifs are together. To achieve this goal, we redefine the first-order proximity and second-order proximity, both of which are different from the definition in \cite{5,6}. The redefined first-order proximity transforms the connectivities in a network from two vertices to more vertices, strengthening intra-motif relationships as connectivity exists between any two vertices in a motif. The redefined second-order proximity captures the relationship between the two vertices which do not reside in a motif but share many redefined neighbors. The two redefined proximities also address the problem that two potentially connected vertices exhibit a weak tie, since we add more neighbors for each vertex by redefining neighbors.\par
Aiming to address the complexity and non-linearity of network structures, we adopt a deep learning model called autoencoders which enjoy high non-linearity. The proposed algorithm first compresses the second-order proximity of vertices to low-dimensional vectors. From the perspective of the first-order proximity, we adjust vectors obtained from the second-order proximity to keep the first-order proximity in the vector space. At the same time, a method of negative sampling based on motifs is proposed to optimize the proposed algorithm.\par
To illustrate the effectiveness of our algorithm, we extensively conduct experiments on multiple types of networks, including social, biological and academic networks. Experimental results show that our algorithm outperforms state-of-the-art embedding-based algorithms by 19\% and traditional similarity-based algorithms by 20\% for link prediction.
In addition, these experiments also show that our algorithm performs better in predicting potential weak ties than
the baseline algorithms.
At last, we investigate the influence of motif types on experiments, which is not discussed before but provides insight for selecting the most appropriate motif type for a network.\par
\textbf{Contributions:} The contributions of this paper are outlined as follows:
\begin{enumerate}
	\item We propose a novel motif-based embedding algorithm for link prediction. This method seamlessly incorporates motifs and deep learning model into link prediction.
	\item We redefine the first-order proximity and the second-order proximity of two vertices by motifs. The two redefined proximities are beneficial for preserving higher-order structures,
	and can be applied to different kinds of network analysis tasks.
	\item Experiments on six networks empirically demonstrate that our algorithm achieves 19\% gain over embedding-based algorithms and 20\% gain over baseline algorithms for link prediction.
\end{enumerate}\par

\textbf{Organization:} The rest of this paper is organized as follows: In Section~\ref{sec2}, we give an overview of related work on network embedding and link prediction. Section~\ref{sec3} introduces some important preliminaries, including the redefined first-order proximity and the second-order proximity. In Section~\ref{sec4}, we present the overall framework of our proposed algorithm and the loss function in detail. In Section~\ref{sec5}, we evaluate the effectiveness of the algorithm over three distinct types of networks. Finally, Section~\ref{sec6} concludes the paper.

\section{Related Work}\label{sec2}
In this section, we review key applications of motifs, and algorithms for network embedding and link prediction. Besides, we state the relationship and difference between our study and some similar studies.
\subsection{Motif}{
	Previously, the concept of motif is used to apply largely to biological networks. Uri Alon \cite{7} discusses two types of transcription network: sensory networks and developmental networks where motifs can be found. Grundy et al. \cite{8} propose motif-based HMMs to solve the problem that HMMs require a relatively large training set. Recently researchers apply motifs to multiple networks and a variety of scenes. Benson et al. \cite{9} propose a motif-cut approach to improve network clustering accuracy, which is successfully applied to social networks, academic networks, and so on. Arenas et al. \cite{10} have used motifs to define which communities vertices belong to. Xia et al. \cite{32} surveyed measures for network motifs.\par
	For detecting motifs in networks, Milo et al. \cite{19} propose an approach to find the striking appearance of motifs in networks. The approach only selects patterns appearing in real networks with the numbers being significantly higher than those in the randomized networks. S. Wernicke proposes an algorithm \cite{42} that can enumerate and sample subgraphs by selecting a random edge and randomly extending the subgraph. Based on the algorithm \cite{42}, Wernicke et al. \cite{20} devise a tool, namely FANMOD, for network motifs detection.
}
\subsection{Network Embedding}{
	\subsubsection{Pair-wise Connectivity}
	Inspired by the success of word vectors, DeepWalk \cite{3} considers a vertex sequence sampled by random walk as a sentence. These sentences can be used to learn a vector representation for each vertex by the power of SkipGram \cite{12}. LINE \cite{5} uses the first-order and the second-order proximity to learn vectors. Meanwhile, edge sampling and negative sampling \cite{13} are used to optimize the model.
	GraRep \cite{14} illustrates the importance of using global information when learning vector representations. Based on the discussion in the study \cite{15} where negative sampling is equal to matrix factorization, GraRep can obtain a low-dimensional vector representation for each vertex using SVD at each step, and concatenate all vectors in \textit{K} step as the last vector representations. SDNE \cite{6} uses a deep learning model that uses the second-order proximity as input to mine highly non-linear relations between vertices and attaches the first-order proximity as a supervised component to the model. The algorithm is successfully applied to link prediction. Since SDNE and LINE only consider pair-wise similarity, the first-order and second-order proximities defined by the similarity can not preserve higher-order structures.\par
	Besides homophily, some networks exhibit structural equivalence or structural similarity. Node2vec \cite{3} defines a second-order random walk using return parameter \textit{p} and in-out parameter \textit{q}, and  also shows effectiveness on link prediction. Different from Node2vec that considers both homophily and structural equivalence of networks, Struc2Vec \cite{16} exclusively focuses on structural identity by vertex sequences ordered by degree of vertices. This approach shows great superiority over Node2vec on node classification.
	\subsubsection{Higher-order Structure}
	The approaches listed above ignore an important characteristic of networks, i.e., higher-order structures. Recently, some researchers have begun to leverage subgraphs, graphlets or motifs to improve the performance of vector representations used to perform network tasks. SNS \cite{17} defines the structural similarity of two vertices based on graphlet and orbit, which can be combined with original methods of learning word2vec, such as CBOW. HONE \cite{18} defines the problem of higher-order network representation learning based on network motifs. This approach defines a new adjacency matrix for each motif from two-node to four-node and demonstrates its effectiveness on link prediction.
	However, this approach does not indicate why motifs are informative of link prediction, and involves a number of matrix multiplications and linear matrix factorizations.
	As a result, HONE has high computational complexity and, as our experiment results show,
	may not able to fully capture complex network structures and relationships between higher-order structures.
}
\subsection{Link Prediction}{
	\subsubsection{Tradititonal Similarity-based Algorithms}
	Traditional similarity-based algorithms \cite{1} for link prediction mainly focus on neighbors of two potentially connected vertices to measure their similarity. Common Neighbors directly simply counts common neighbors. Jaccard's Coefficient divides the number of common neighbors by the number of all neighbors of the two vertices. Adamic-Adar also counts common neighbors, but assigns the less-connected neighbors more weight. Preferential attachment assumes that two vertices
	of higher degrees are more likely to link to each other.
	
	\subsubsection{Embedding-based Algorithms}
	Embedding-based algorithms measure the similarity of two vertices by their learned vectorized features. SDNE \cite{6} uses the cosine as an indicator of similarity. Node2vec \cite{3} designs four different operators over vertices' features to generate edges' features, including Average, Hadamard, Weighted-L1 and Weighted-L2. HONE \cite{18} uses Average operation to construct edges' features. DeepGL \cite{rossi2018deep} uses the four operators to construct edges' features for inductive network representation. The paper \cite{3} illustrates that embedding-based algorithms can outperform traditional similarity-based algorithms on link prediction. Wang et al. \cite{wang2017predictive} propose a predictive network representation learning named PNRL to solve the structural link prediction problem. PNRL jointly optimizes two objectives for observed links and assumed hidden links. Tu et al. \cite{tu2018unified} propose a Community-enhanced
	Network Representation Learning named CNRL. The algorithm shows its superiority on link prediction. Liao et al.  \cite{liao2018attributed} propose a network embedding for attributed social networks, and show that the method achieves substantial gains on the task of link prediction. Zou et al. \cite{zou2019encoding} propose an encoder-decoder model for graph generation. The encoder is a Gaussianized graph scattering transform and the decoder can be adapted to link prediction. RGCN \cite{schlichtkrull2018modeling} uses graph convolutional networks to learn embedding vectors and uses a tensor factorization model to predict links.
}
\subsection{Similar Work}{
	There are some studies similar to our. Both E-LSTM-D \cite{chen2019e} and MODEL aim at predicting links in networks and use auto-encoder. However, E-LSTM-D focuses on dynamic networks and thus needs multiple graphs. MODEL focuses on static networks and thus needs only one graph. S-LSTM-D combines auto-encoder and a stacked long short-term memory (LSTM), a deep learning model, and does not consider the higher-order structure. MODEL combines auto-encoder and motifs, higher-order structures that naturally exist in networks. Qi et al. propose a concept of subgraph network (SGN) that can be applied to network models \cite{xuan2019subgraph}. Both SGN and MODEL consider subgraphs, but SGN uses various types of subgraphs. MODEL uses only one type of subgraphs, i.e., motif. SGN constructs structural features without considering the nonlinear network structure, while MODEL learns features by auto-encoder that addresses the nonlinear network structure. The goal of the paper \cite{xuan2019subgraph} seams to address network classification; our goal is to predict links of networks. Both the study \cite{fu2018link} and MODEL leverage network property. But the study uses the centrality features; MODEL uses motifs. The study combines network property with line graphs to predict link weights, i.e., to determine the weight of links, while MODEL combines network property with auto-encoder to predict links, i.e., to judge whether there is a link between two vertices. Both RUM \cite{yu2019rum} and MODEL leverage motifs to capture high-order network structure. However, RUM needs to be combined with existing embedding algorithms, such as DeepWalk and LINE, and does not consider the nonlinear network structure. On the contrary, MODEL is an individual algorithm and uses auto-encoder to capture the nonlinear network structure. Motif2Vec \cite{dareddy2019motif2vec} also leverages motifs, but its goal is to design a new random-walk method, which is unsatisfactory to capture the nonlinearity of the network structure. Furthermore, Motif2Vec focuses on heterogeneous networks, while MODEL focuses on homogeneous networks. The study \cite{abuoda2019link} defines topological features by motifs from three-node to five-node, without considering the nonlinearity of the network structure. MODEL learns features only by one type of motifs and considers the nonlinearity. The study \cite{abuoda2019link} sees link prediction as a binary classification problem, while MODEL sees it as a probability problem.
}

\section{Preliminaries}\label{sec3}
{
	In this section, we list some important preliminaries and notations for this paper. It is worthwhile to note that the definitions of the first-order proximity and the second-order proximity in this section are redefined by motifs, different from the definitions from the papers \cite{5,6}. Some key notations are shown in Table  \ref{table:1}.\par
	\begin{table}[htbp]
		\centering
		\caption{Key notations in this paper}\label{table:1}
		\begin{tabular}{c|l}
			\hline
			\hline
			Notations & Meaning\\[1pt]
			\hline
			\hline
			\textit{G} & A graph\\[1pt]
			\textit{V} & The vertex set\\[1pt]
			$|V|$ & The number of vertices\\[1pt]
			\textit{v}$_{i}$ & The vertex \textit{i}\\[1pt]
			\textit{E} & The edge set\\[1pt]
			$|E|$ & The number of edges\\[1pt]
			\textit{M} & A motif/A motif type\\[1pt]
			\{\textit{M}\} & The motif set\\[1pt]
			\textit{M}$_{\textit{xy}}$ & A motif of order \textit{x} and number\textit{ y}\\[1pt]
			\textit{d} & The dimention of vector representations\\[1pt]
			\textit{y}$_{\textit{i}}$ & The vector representation of vertex \textit{i}\\[1pt]
			\textit{N}$_{\textit{i}}$ & The first-order proximity of vertex \textit{i}  and other vertices\\[1pt]
			$N(i)$ & The set of neighbors of vertex \textit{i}\\[1pt]
			\textit{f},\textit{g} & Mapping functions \\[1pt]
			$x$ & An input vector of the model\\[1pt]
			$x^j$ & The $j_{th}$ element of $x$\\[1pt]
			$y^k$ & An output vector of $k_{th}$ hidden layer in encoders/decoders\\[1pt]
			$x'$ & An ouput vector of decoders\\[1pt]
			$z$ & The penalty vector\\[1pt]
			$W_k$ & The weights of $k_{th}$ hidden layer in encoders\\[1pt]
			$B_k$ & The biases of $k_{th}$ hidden layer in encoders\\[1pt]
			$W_k'$ & The weights of $k_{th}$ hidden layer in decoders\\[1pt]
			$B_k'$ & The biases of $k_{th}$ hidden layer in decoders\\[1pt]
			$K$ & The number of hidden layers in encoders/decoders\\[1pt]
			$\theta$ & The set of parameters to be learned\\[1pt]
			$\odot$ & The Hadamard product of matrices\\[1pt]
			$\| \; \|_F $ &  Frobenius norm\\[1pt]
			\hline
		\end{tabular}
	\end{table}\par
	\subsection{Graph and Motif}
	We denote a graph as $G=(V,E)$, where \textit{V} denotes the set of all vertices and \textit{E} denotes the set of all edges in \textit{G}, i.e., $(v_{i},v_{j})\in E $ if \textit{v}$_{\textit{i}}$ and \textit{v}$_{\textit{j}}$ are connected.
	A motif is defined as a connected subgraph with a small number of vertices. Formally, we denote a motif as $M = (V_M,E_M)$. If a motif is a connected subgraph from graph \textit{G}, $(v_i,v_j) \in E$ for any $(v_i,v_j) \in E_M$. A motif is not merely any subgraph. It exists in a type of networks significantly more frequently than a random subgraph, reflecting an underlying process specific to the type of networks.  For example, feed-forward loops are known to be prevalent and has functional significance in Protein–protein interaction networks.
	
In general,	network motif search is NP-hard. Higher-order motifs require more time to find and  more space to store. Aiming at reducing the time complexity and space complexity, we only consider all three-node and four-node motifs in this paper. Besides, our proposed approach can also be extended to arbitrary-node motifs. In Fig. \ref{figure:2}, we draw all types of three-node and four-node motifs. Note that a motif $M_{xy}$ represents the type of motif with order \textit{x} and number \textit{y}.
	Efficient algorithms are available for finding motifs in a network \cite{20,21}.

	\begin{figure}[htbp]
		\centering		
		\includegraphics[scale=0.6]{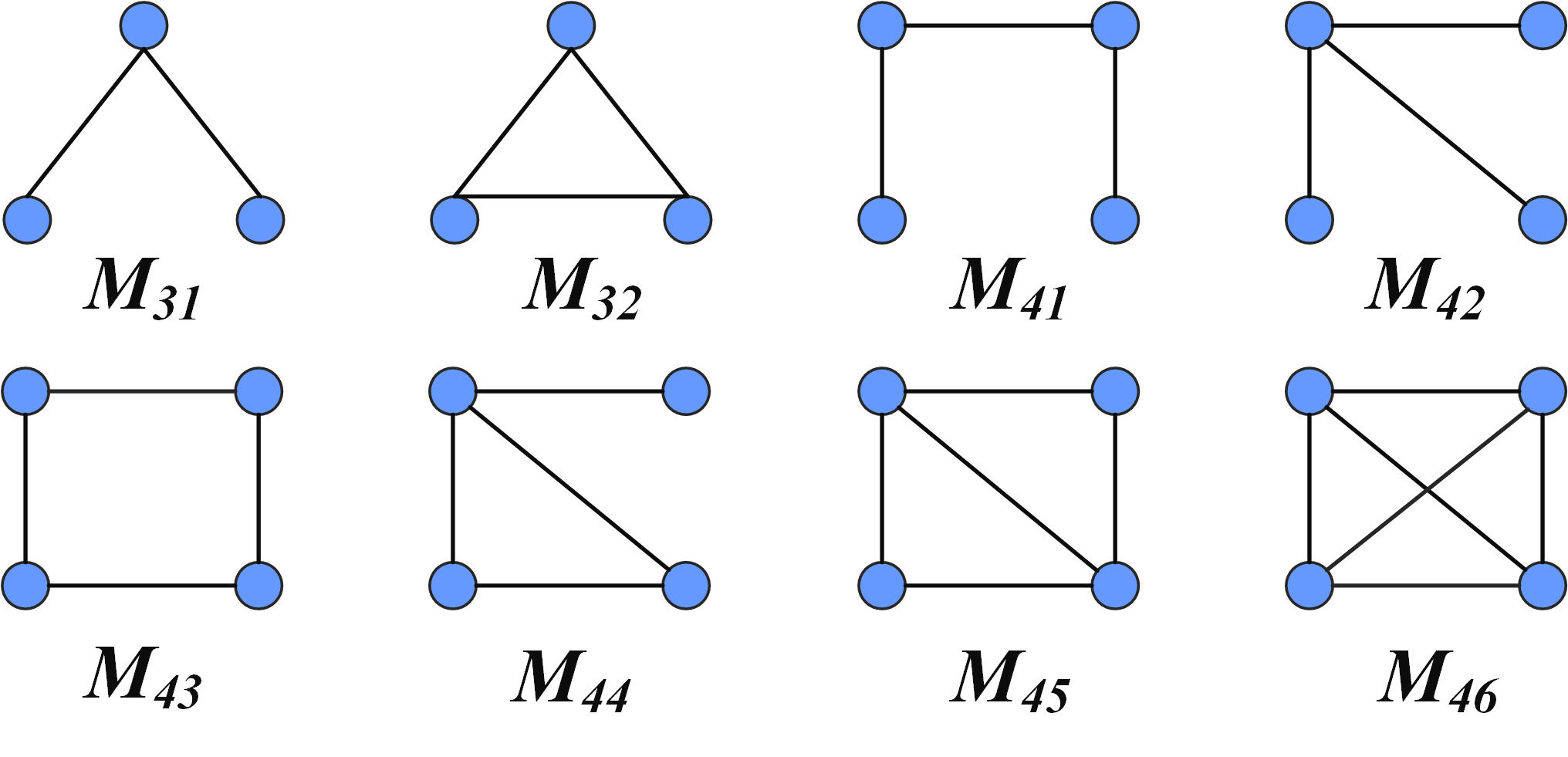}
		\caption{All three-node and four-node motifs. Circles represent vertices, and lines represent edges.}\label{figure:2}
	\end{figure}\par
	\subsection{First-order Proximity and Second-order Proximity}
	\textbf{First-order proximity: } The first-order proximity is directly defined on pair-wise vertices. If $v_{i}$ and $v_{j}$ reside at the same motif, there exists the first-order proximity between $v_{i}$ and $v_{j}$. Otherwise, the proximity is 0. The more motifs both $v_{i}$ and $v_{j}$ reside at are, the larger the first-order proximity between $v_{i}$ and $v_{j}$ is.\par
	The way to define the first-order proximity is different from the definition in LINE \cite{5} and SDNE \cite{6}. LINE and SDNE define the  first-order proximity on two connected vertices, while we define it on any two vertices in a motif, despite whether they are connected. Taking $M_{43}$ in Fig. \ref{figure:2} as an example, we attach the first-order proximity on any two diagonal vertices and any two connected vertices as they all reside at $M_{43}$. However, LINE and SDNE do not attach the first-order proximity on two diagonal vertices since they are not connected. It is worth noting that our way defining the proximity covers the definition in LINE and SDNE. If we only consider two-node motifs, the two definitions are equivalent. Since the first-order proximity exists between any two vertices in a motif, the relationship of intra-motif can be strengthened.\par
	In this paper, let $N_i=\{w_{i1},w_{i2},\dots,w_{in}\}$ denote the first-order proximity of $v_i$ and other vertices. $w_{ij} \in R\;(R>=0)$ denotes the first-order proximity of $v_i$ and $v_j$, which is defined as the number of motifs where $v_i$ and $v_j$ occur simultaneously. \\
	\textbf{Second-order proximity:} The second-order proximity is defined on neighbors of two vertices. It is founded on the cognition that the more common neighbors two vertices have, the more similar the two vertices are \cite{5,6}.
	In this paper, we redefine the neighbors of every two nodes via a specific motif.
	\par
	According to the first-order proximity we defined, the second-order proximity of $v_i$ and $v_j$ can be calculated by the similarity of $N_i$ and $N_j$. If $v_i$ and $v_j$ share a great number of same neighbors, the two nodes and their neighbors will be close in the vector space, which leads to the two motifs of $v_i$ and $v_j$ become close. \par
	\subsection{Autoencoder}
	Autoencoder, introduced first in \cite{27} by Hinton et al., is a non-linear way of reducing the data dimension by multi-layer neural network.\par
	A universal framework of autoencoder is shown in Fig. \ref{figure:3}. An autoencoder consists of two parts,  one part compressing high-dimensional data \textit{x} to low-dimensional data \textit{y} and the other part decompressing \textit{y} to high-dimensional data $x'$ seen as reconstructed \textit{x}. Each part is actually a multi-layer neural network with a large number of non-linear activation functions.
	We use $f_\theta$ and $g_{\theta'}$ to denote the encoding and decoding function respectively, thus the process of encoding and decoding is equal to address the function $\arg\min_{\theta,\theta'}\ell(x,g_{\theta'}(f_\theta(x)))$, wher $\ell(x,x')$ is the function measuring the difference of $x$ and $x'$, e.g., Euclidean distance.
	\begin{figure}[htbp]
		\centering		
		\includegraphics[scale=0.3]{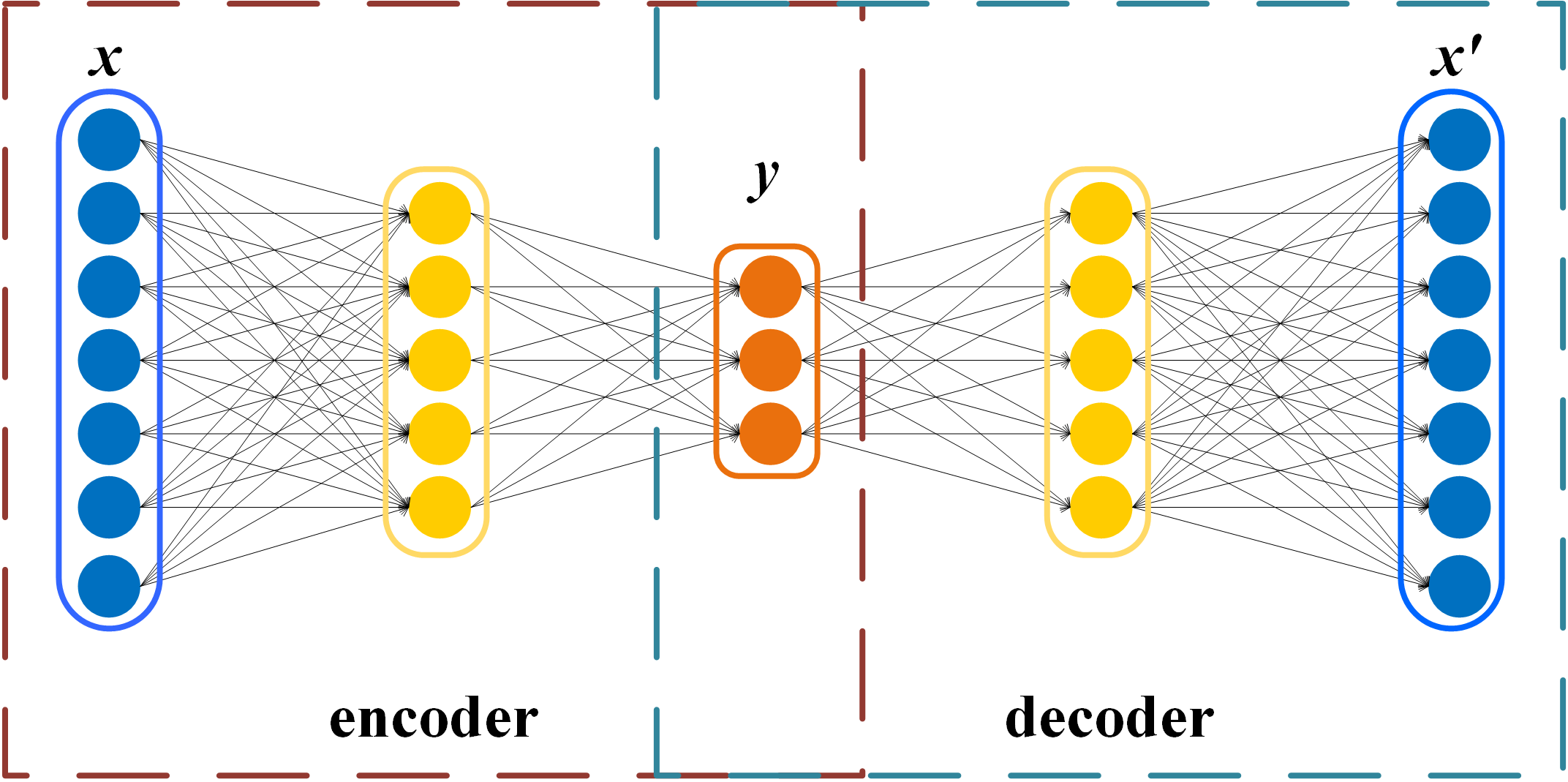}
		\caption{A universal framework of autoencoder. The left part is an encoder and the right part is a decoder. In the autoencoder, $x$ is the input vector, $y$ is the output vector,
			and  $x'$ is the reconstructed vector of $x$.
		}\label{figure:3}
	\end{figure}
	\subsection{Problem Definition}
	
	\textbf{Motif-based Embedding:} Given all motifs with a specific type in a network, we need to learn a mapping function $f: V \rightarrow y \in R^d$, where $d \ll n$. We hope that learned mapping function can preserve well higher-order structures in original networks, i.e., the similarity of intra-motif and relationships between motifs.\\
	
	\textbf{Link prediction:} Link prediction aims to predict missing edges or edges that will appear in the future. 
	A graph $G=(V,E)$ can be regarded as $G=\left\{(v_i, v_j)\in E\right\}\subseteq V \times V$, a subset of all possible edges.
	In practice, we may not be able to model every edge, particular those considered weak ties.
	
	The objective of \emph{link prediction} is to find a scoring function $S: V\times V \mapsto [0, 1]$ such that 
	$S(v_i, v_j)$ is high if $(v_i, v_j)\in E$ and $S(v_i, v_j)$ is low if $(v_i, v_j)\not \in E$. That is, we aim to maximize
	$$\sum_{(v_i, v_j)\in E}\log S(v_i, v_j) + \sum_{(v_i, v_j)\not\in E}\log (1 - S(v_i, v_j)).$$

	Since we use a \textit{d}-dimensional vector to represent each vertex in this paper, we can use the cosine similarity of $\boldsymbol{y}_i$ and $\boldsymbol{y}_j$ to indicate the score that there is an underlying edge between $v_i$ and $v_j$. That is 
	$$S(v_i, v_j) = \cos(\boldsymbol{y}_i, \boldsymbol{y}_j).$$
	
}

\section{Design of MODEL}\label{sec4}
{
	\begin{figure*}[t]
		\centering		
		\includegraphics[scale=0.6]{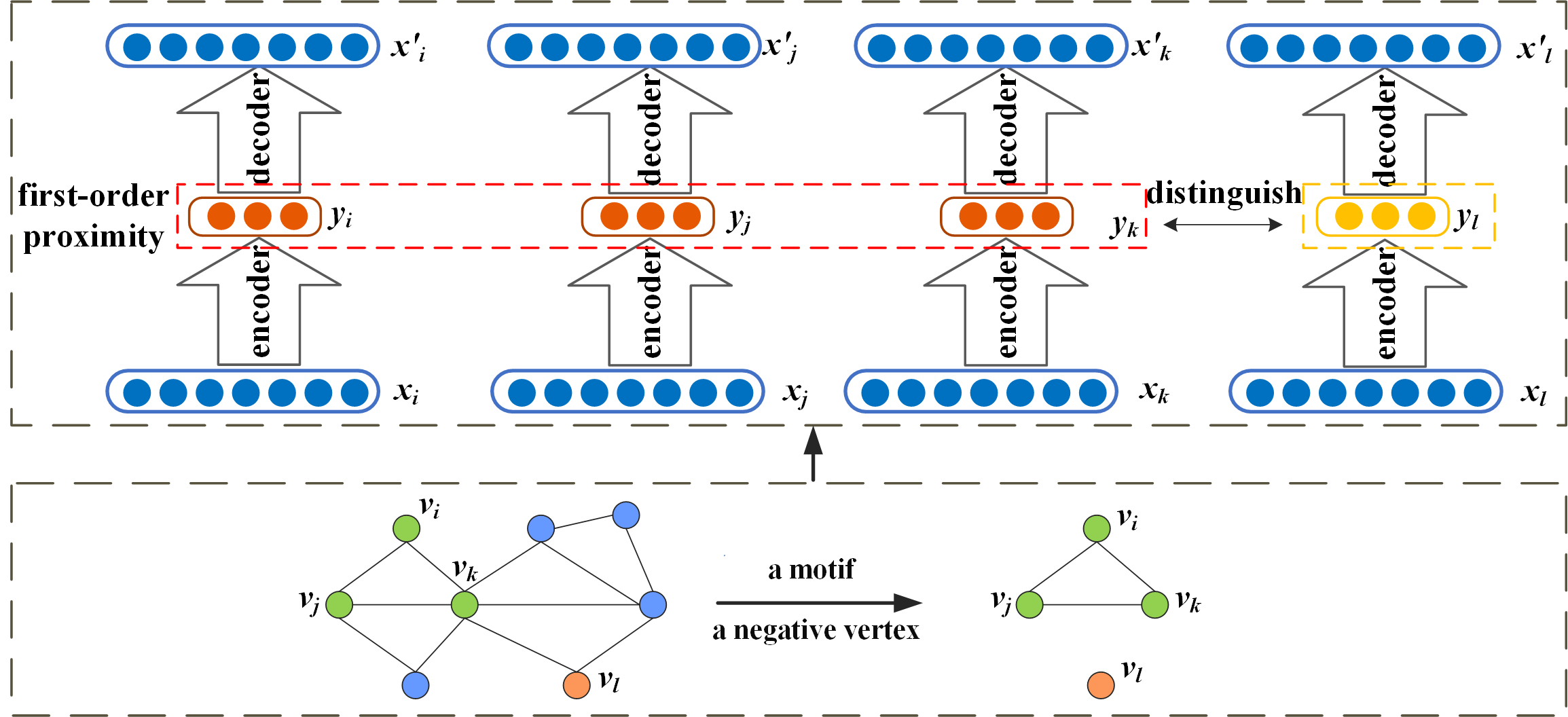}
		\caption{The overall framework of MODEL. The model first selects a motif composed of $v_i$, $v_j$ and $v_k$,  and a negative vertex $v_l$. The vectorized neighbors of the four vertices are inputted simultaneously into four autoencoders sharing the same parameters. The outputs of autoencoders, $y_i$, $y_j$, $y_k$ and $y_l$, serve as learned features of their corresponding vertices.
			Due to the first-order proximity, a loss term is added for $y_i$, $y_j$ and $y_k$.
			Meanwhile, a separate loss term is used to distinguish a motif from a negative vertex in the vector space.}\label{figure:4}
	\end{figure*}\par
	In this section, we first give the overall framework of proposed algorithm MODEL. Subsequently, We introduce how to preserve higher-order structures, i.e., the second-order proximity and the first-order proximity, and how to optimize them by sampling negative nodes. Last, we conclude the complexity of training the model.
	\subsection{Framework}{
		To learn a mapping that preserves higher-order structures in networks,
		we perform the next two steps: First, we construct a vector representation for each vertex by their neighbors. A simple way is that we directly use $N_i$ as a vector to represent $v_i$. However, due to its high sparsity, discrete values and the large dimension, the vector is inappropriate for downstream network tasks. In practice, we can compress $N_i$ to a low-dimensional vector with entries being real values to address the problem. In the process of compression, we employ an autoencoder that enjoys non-linearity and flexibility \cite{6,22,23}. Second, considering the first-order proximity, we need to attach a loss term on vector representations of all vertices in a motif.
		An effective and widely used way is to minimize the Euclidean distances of vectors of any two vertices in a motif. For taking better advantage of the power of the first-order proximity, neighbors of all vertices in a motif are inputted to encoder simultaneously, so that vectors of the vertices can be obtained at the same time.\par
		
		We show the framework of the proposed approach in Fig. \ref{figure:4}. The model consists of four autoencoders which share the same weights, biases and activation functions.
		The part below is encoder and
		the one above is decoder.
		Given a motif composed of $v_i$, $v_j$ and $v_k$, the model first input neighbor vectors $x_i$, $x_j$ and $x_k$, corresponding to $v_i$, $v_j$ and $v_k$ respectively. By encoding and decoding, $y_i$, $y_j$ and $y_k$ are obtained as vector representations, which can reflect the second-order proximity between these vertices. Due to the first-order proximity in a motif, a loss term is attached to regularise vector representations. To optimize the model, an autoencoder for a negative vertex $v_l$ is added. We hope the model can distinguish $y_i$, $y_j$ and $y_k$ from $y_l$ in the vector space. \par
		
		The Model in Fig. \ref{figure:4} is drawn based on three-node motifs, so that there are only four auto-encoders, in which $v_i$, $v_j$, and $v_k$ compose a three-node motif and $v_l$ is a negative vertex. Even so, the model can be transformed easily to be of four-node motifs and even higher-order motifs. For example, we can add another autoencoder sharing the same parameters, so that the new model can be applied to four-node motifs. All next introductions about the proposed algorithm are based on three-node motifs. Meanwhile, we list some key symbols of the model in Table   \ref{table:1}. Note that a vector with subscript $i$ represents the specific vector to $v_i$, e.g., $x_i$ is the input vector of $v_i$.\par
		Next, we detail how our model learns a vector representation for each vertex to preserve the second-order proximity and the first-order proximity, and how to add noise to optimize the model.
	}
	\subsection{Preserving Second-order Proximity}{
		For preserving the second-order proximity, we first need to obtain neighbors of each vertex according to Section \ref{sec3}. These neighbors are stored in $N(i)$ and seen as the input vector $x_i$ of $v_i$ in encoders. Through Equation \ref{eq1} and \ref{eq2}, where $\sigma$ denotes the activation function, such as \textit{sigmoid}, we can obtain output vectors of encoders as vector representations. The more similar two input vectors are, the more similar two output vectors are.
		\begin{equation}\label{eq1}
		y_i^1=\sigma(W_1x_i+B_i)
		\end{equation}
		\begin{equation}\label{eq2}
		y_i^k=\sigma(W_ky_i^{k-1}+B_k)\quad k=2,\dots,K
		\end{equation}
		\par
		After compression, decoders accept output vectors of encoders as input vectors and then reconstruct input vectors of encoders. Through Equation~\ref{eq3} and \ref{eq4}, we can obtain reconstructed vector $x_i'$ of vector $x_i$. In Equation~\ref{eq3}, $y_i^0$ is the input vector of decoders and also the output vector of encoders.
		\begin{equation}\label{eq3}
		y_i^k=\sigma(W_k' y_i^{k-1}+B_k')\quad k=1,\dots,K-1
		\end{equation}
		\begin{equation}\label{eq4}
		x_i'=y_i^K=\sigma(W_K' y_i^{K-1}+B_K')
		\end{equation}
		\par
		In the process of encoding and decoding, we minimize the distances of original vectors and reconstructed vectors, i.e., $\|x_i-x_i'\|_2^2$ for input vector $x_i$. Given all motifs denoted by $\{M\}$ in a network, the loss function is Equation \ref{eq5}, where $\| \; \|_2$ denotes the Euclidean norm.
		\begin{equation}\label{eq5}
		\ell_{2nd}=\sum_{M \in \{M\}}\sum_{i \in M}\|x_i-x_i'\|_2^2
		\end{equation}\par
		Due to the high sparsity in a lot of networks, a large number of elements in \textit{x} are zero,  so that autoencoders are prone to reconstruct these elements with value zero \cite{6}. To address this problem, we add a penalty vector $z_i$ for input vector $x_i$ in Equation \ref{eq5}. If $x_i^j >$ 0, $z_i^j = \beta\,(\beta>1)$, else $z_i^j =1$. Through Equation \ref{eq6} that gives more weight to non-zero elements than zero elements, autoencoders are prone to reconstruct non-zero elements of input vector \textit{x}.
		\begin{equation}\label{eq6}
		\ell_{2nd}=\sum_{M \in \{M\}}\sum_{i \in M}\|(x_i-x_i')\odot z_i\|_2^2
		\end{equation}
		In the above equation, $\odot$ denotes the Hadamard product which represents element-wise multiplications for matrices.
	}
	\subsection{Preserving First-order Proximity}{
		Besides the second-order proximity, we need to capture the intra-motif similarities, i.e., the first-order proximity between vertices. Given a motif consisting of $v_i$, $v_j$, and $v_k$, we can obtain vector representations $y_i$, $y_j$, and $y_k$  by autoencoders. Because of the existence of the first-order proximity between any two vertices in $v_i$, $v_j$, and $v_k$, we need to minimize the Euclidean distances of vector representations of any two vertices. The equation is:
		\begin{equation}\label{eq7}
		\ell=\| y_i-y_j\|_2^2+\| y_i-y_k\|_2^2+\| y_j-y_k\|_2^2
		\end{equation}\par
		Through the above equation, vector representations of all vertices residing in the same motif
		are close to each other. Note that for three-node motifs, there are three Euclidean distances in Equation \ref{eq7}. If we consider four-node motifs, there will be six Euclidean distances. Given all motifs denoted by $\{M\}$ in a network, the loss function is Equation \ref{eq8}.
		\begin{equation}\label{eq8}
		\ell_{1st}=\sum_{M \in \{M\}}\sum_{i,j \in M,i\neq j}\|y_i-y_j\|_2^2
		\end{equation}\par
		From the above equation, we can find that the more the number of motifs both $v_i$ and $v_j$ reside in is, the closer $y_i$ and $y_j$ are.
	}
	\subsection{Negative Sampling}{
		In Section 4.3, we only strengthen intra-motif relationships. As a result, the algorithm may converge to the solution that all motif-wise relationships are similar and cannot distinguish a motif from other motifs, which harms experimental performance. In the study of \cite{13}, an effective approach called negative sampling is proposed. The main idea of this approach is to distinguish right from noises. Inspired by this approach, we hope to distinguish what vertices can compose a motif from a noise. The noise can be sampled randomly or based on the number of motifs each vertex resides in. The distinguishing process can be formulated as follows:
		{\small 	\begin{equation}\label{eq9}
			\ell_{dis}=\text{max}(\lambda+\sum_{i,j \in M,i\neq j}\|y_i-y_j\|_2^2-\mu\sum_{i \in M}\|y_i-y_l\|_2^2,0)
			\end{equation}}
		where we denote a noise by $v_l$. A balance factor $\mu$ is added in Equation \ref{eq9}, since the number of Euclidean distances in the first integral sum may be greater than that in the second integral sum. For three-node motifs, $\mu$ is one while it is 3/2 for four-node motifs.
		$\lambda$ is a slack parameter that represents the margin by which
		the second sum should exceed the first sum.
		
		Combined with Equation \ref{eq9}, the Equation \ref{eq8} is transformed to Equation \ref{eq10}:
		\begin{equation}\label{eq10}
		\ell_{1st}=\sum_{M \in \{M\}}\ell_{dis}
		\end{equation}	
	}
	\subsection{Optimization}{
		From Section 4.2 and 4.3, we obtain two loss functions to preserve the first-order proximity and the second-order proximity respectively. Meanwhile, to prevent overfitting, a loss function of the regularization on weights of autoencoders needs to be added\cite{6}. The loss function is Equation \ref{eq11}:
		\begin{equation}\label{eq11}
		\ell_{reg}=\sum_{k=1}^K(\| W_k \|_F^2+\| W_k' \|_F^2)
		\end{equation}
		where $\| \; \|_F $ denotes the Frobenius norm.\par
		Combining Equation \ref{eq5}, \ref{eq10} and \ref{eq11}, we can obtain a joint loss function as follows:
		\begin{equation}\label{eq12}
		\ell_{loss}=\ell_{2nd}+\alpha\ell_{1st}+\gamma\ell_{reg}
		\end{equation}
		where $\alpha$ and $\gamma$ are the weights of the first-order proximity and the regularization respectively.\par
		To minimize Equation \ref{eq12} and make the algorithm converge fast, we use Adam algorithm \cite{24} and back-propagation \cite{25} to optimize our model. We summarize the process of our approach in Algorithm 1.\par
		In Algorithm 1, we first need to look for all motifs $\{M\}$ with type \textit{M} and then obtain neighbors for each vertex according to Section \ref{sec3}. At step 3, to prevent the model suffering from local optimality, we use DBN \cite{26} to pre-train parameters $\theta$ at first.
	}
	\subsection{Complexity Analysis}{
		When looking for all motifs, we resort to FANMOD \cite{20} which is easy to implement and can quickly enumerate all motifs.\par
		In the process of training our model, the data used is a small batch of motifs. Therefore, the time complexity is $O(c\Delta dTK)$ where \textit{c} is the order of a motif. Since only the three-node and four-node motifs are considered, \textit{c} is either 3 or 4. \textit{T} is the maximal iteration number, and is set as 200 in all experiments. $d$ is the dimension of vector representations, and $K$ is the number of layers in encoders or decoders. At last, it is easy to find that the time complexity is linear to the number of motifs. During experiments, only a small portion of all motifs in a network is needed, which avoids the problem that some networks contain tens of thousands of motifs.
	}
	\begin{table}[htbp]
		\centering
		\begin{tabular}{l}
			\hline
			\textbf{Algorithm 1: MODEL}\\[1pt]
			\hline
			\textbf{Input:} a network $G=(V,E)$, motif type \textit{M},
			parameters $\alpha,\gamma$,\\[1pt] $\beta,\lambda$, batch size $\Delta$, learning rate $\delta$\\[1pt]
			\textbf{output:} vector representations y of all vertices\\[1pt]
			1: Look for all motifs $\{M\}$ with type \textit{M} in \textit{G}\\[1pt]
			2: Obtain redefined neighbors for each vertex in \textit{G}\\[1pt]
			3: Initiate parameters $\theta=\{W_k,B_k,W_k',B_k'\}_{k=1}^K$\\[1pt]
			4: \textbf{Repeat:}\\[1pt]
			5: \quad  Randomly select a subset $\{M\}_\Delta$ with size $\Delta$ from $\{M\}$\\[1pt]
			6: \quad Calculate $\ell_{loss}$ according to Equation \ref{eq12}\\[1pt]
			7: \quad Update parameters $\theta$ by back-propagation and \\[1pt] \qquad Adam with the learning rate $\delta$\\[1pt]
			8:\textbf{ Until convergence}\\[1pt]
			\hline
		\end{tabular}
	\end{table}
}

\section{Experiments and Analysis}\label{sec5}
{
	\subsection{Datasets}
	In this paper, we select six networks with three distinct types to comprehensively evaluate the effectiveness of our proposed algorithm. The six networks are introduced as follows.\par
	Youtube \cite{28} and LiveJournal \cite{28} are social networks where vertices denote users and each edge indicates the friendship between corresponding two vertices. In the two networks, we randomly select 5,346 vertices in Youtube and 2,456 vertices in LiveJournal as our experimental datasets and ensure that there are not isolated vertices in the two datasets.\par
	Bio-sc-cc \cite{29} and Bio-sc-ht \cite{29} are biological networks where vertices denote genes and each edge indicates the interaction between corresponding two genes. Bio-sc-cc has 2,223 vertices and 34,879 edges, and Bio-sc-ht has 2,084 vertices and 63,027 edges.\par
	DBLP \cite{28} and Ca-GrQc \cite{30} are academic networks where vertices denote scholars and each edge indicates the collaboration between the corresponding two scholars. We randomly select 4,424 vertices from the original DBLP network as our experimental dataset and ensure that there are not isolated vertices in the dataset. Ca-GrQc has 4,158 vertices and 13,422 edges.\par
	Some detailed statistics about the six network datasets are summarized in Table \ref{table:3}.\par
	\begin{table}[htbp]
		\centering
		\caption{Statistics of the six datasets}\label{table:3}
		\begin{tabular}{c|c|c|c|c}
			\toprule
			datasets & $|V|$ & $|E|$ & Avg. Degree & Network Type\\
			\midrule
			Youtube & 5,246 & 24,121 & 9.02 & Social\\
			LiveJournal & 2,456 & 188,490 & 153.49 & Social\\
			Bio-sc-cc & 2,236 & 34,879 & 31.2 & Biological\\
			Bio-sc-ht & 2,084 & 63,027 & 60.49 & Biological \\
			DBLP & 4,424 & 12,169 & 5.5 & Academic \\
			Ca-GrQc & 4,158 & 13,422 & 6.46 & Academic\\
			\bottomrule
		\end{tabular}
	\end{table}
	\subsection{Baseline Algorithms}

	Although there are a large number of algorithms on network embedding, only small of them illuminate their effectiveness for link prediction. In this paper, we use the following algorithms as baselines, including three traditional methods that can achieve good performance in link prediction \cite{1}.\par
	DeepWalk \cite{3}: It randomly samples many vertex sequences via random walk and generates vector representations of vertices using Skip-Gram on these vertex sequences.\par
	Node2vec \cite{4}: It can be seen as the extended DeepWalk and randomly samples vertex sequences via $2^{nd}$ order random walk with two parameters \textit{p} and \textit{q}.\par
	SDNE \cite{6}: It employs an autoencoder as a deep model to generate vector representations of vertices. The autoencoder inputs the adjacent vector of each vertex and adds the first-order proximity as supervised information.\par
	HONE \cite{18}: It considers all types of motifs from two-node to four-node, and generates a local \textit{k}-step embedding for each motif. At last, it concatenates all local embeddings for all motifs in \textit{k} steps, and obtains low-dimensional vector representations via matrix factorization. \par
	SCAT \cite{zou2019encoding}: It uses a Gaussianized graph scattering transform as an encoder and a fully connected network as a decoder. The decoder is adapted to link prediction. \par
	RGCN \cite{schlichtkrull2018modeling}: It produces latent feature representations by an encoder and predicts label edges by a decoder that is a tensor factorization model exploiting these representations.\par
	Traditional methods \cite{1}: Common Neighbors (CN) defines the similarity of vertex $i$ and $j$ as $|N(i)\bigcap N(j)|$. Jaccard's Coefficient (JC) defines them as $\frac{|N(i)\bigcap N(j)|}{|N(i)\bigcup N(j)|}$. And Adamic-Adar (AA) Score defines them as $\sum_{t \in N(i)\bigcap N(j)}\frac{1}{\log|N(t)|}$.
	\subsection{Evaluation Metrics}
	For link prediction, precisionK, AUC, and Avg. Rank are three widely used evaluation metrics \cite{1}. In our experiments, we use these metrics to evaluate the performance of our proposed algorithm and other algorithms. The detailed introduction about the three metrics is as follows:
	Let $D^+$ denote the set of positive examples, $D^-$ denote the set of negative examples, and $P(e)(e\in D^+ \cup D^-, D^+ \cap D^-=\varnothing)$ denote the probability that $e \in D^+$. Based on $P$, a decreasing ordered list, denoted by RANK, can be obtained. $RANK(e)$, from 1 to $|D^+|+|D^-|$, represents the rank of \textit{e}. We hope elements in $D^+$ can be ranked in front of these elements in $D^-$.\par
	PrecisionK is the proportion of elements from $D^+$ in the top-\textit{K} elements. It is calculated as follows:
	\begin{equation}
	\text{PrecisionK=}\frac{|\{e|RANK(e)\leqslant \text{K},e \in D^+ \}|}{\text{K}}
	\end{equation}\par
	AUC is related to the ranking quality of all examples. It is calculated as follows:
	\begin{multline}
	\text{AUC}=1-\frac{1}{|D^+||D^-|}\sum_{e^+ \in D^+}\sum_{e^- \in D^-}(\dagger(RANK(e^+)< \\
	RANK(e^-))+0.5\dagger(RANK(e^+)=RANK(e^-)))\label{eq14}
	\end{multline}
	
	where $\dagger(x)$ is 1 if \textit{x} is true, else it equals to 0. From the equation \ref{eq14}, we can know that AUC always exceeds 0.5 or equals to 0.5.
	
	Avg. Rank is the average rank of elements in $D^+$. It is calculated as follows:
	\begin{equation}
		Avg.Rank=\frac{\sum_{e\in D^+}RANK(e)}{|D^+|}
	\end{equation}
	\subsection{Experimental Settings}
	Before doing experiments, we first randomly hide some edges truly existing in original networks as $D^+$ and ensure that the degree of each vertex is greater than zero. In the meantime, we randomly select some edges not existing as $D^-$ and ensure that $|D^+|=|D^-|$. During the process of training, hidden edges are not used. For sparse networks, 20\% edges are hidden in DBLP and 30\% edges are hidden in Youtube and Ca-GrQc. For dense networks, 60\% edges are hidden in Bio-sc-cc, 80\% edges are hidden in Bio-sc-ht and 90\% edges are hidden in LiveJournal. In experiments, we need to predict which edges are from $D^+$ in $D^+ \cup D^-$.\par
	We compare our proposed algorithm with those baseline algorithms introduced in Section 5.2. For embedding based algorithms, we set all embedding dimension to 128. For DeepWalk and Node2vec, we set \textit{walks per node r}=10, \textit{walk length l}=80, and \textit{context size k}=10. The optimal \textit{p} and \textit{q} are selected from \{0.25, 0.5, 1, 2, 4\} for Node2vec. For HONE, we also consider all motifs with 2-4 vertices and select step \textit{K} from \{1, 2, 3, 4\}. According to \cite{6} and \cite{zou2019encoding}, we use two-layers encoders for SDNE and SCAT. The dimensions of hidden layers are set to 1000 and 512, respectively. For RGCN, we use two-layers graph convolutional networks with 512 and 128 hidden units, and regard existent and nonexistent edges as two types of edges. In the studies \cite{18} and \cite{zou2019encoding}, there are four variants of HONE and two variants of SCAT. In experiments, we only show the best performance in these variants for the comparison.\par
	For our proposed algorithm, different types of motif have different performance. We conduct experiments over all motif types from $M_{31}$ to $M_{46}$ and show the best performance in them. In section 5.6, we show the influence of motif types on experiments. We select a negative vertex based on the number of motifs it resides in. The hyper-parameters $\alpha$, $\beta$, $\lambda$ and $\gamma$ are set 20, 30, 30 and 0.0001 respectively. The learning rate is set 0.001 and the size of a batch is set 500. All encoders are single-layer neural networks. We select $tanh$ as the activation function, because we hope that the final vector representations contain both positive and negative values. All experiments run 10 times and these results are averaged.\par
	After obtaining the vector representation of each vertex, the cosine value of vectors of two vertices represents the similarity of them. A good algorithm for link prediction can rank elements in $D^+$ in front of elements in $D^-$ based on the similarity.
	
		\begin{table*}[t]
		\centering
		\caption{Comparisons of ten algorithms on AUC}\label{table:4}
		\begin{tabular}{c|c|c|c|c|c|c}
			\toprule
			Algorithm & Youtube & LiveJournal & Bio-sc-cc & Bio-sc-ht & DBLP & Ca-GrQc \\
			\midrule
			MODEL & \textbf{0.843} & 0.919 & \textbf{0.793} & \textbf{0.841} & \textbf{0.944} & 0.820\\
			DeepWalk & 0.684 & 0.724 & 0.677 & 0.746 & 0.901 & 0.761\\
			Node2vec & 0.699 & 0.736 & 0.685 & 0.747 & 0.922 & 0.781\\
			SDNE & 0.662 & 0.747 & 0.692 & 0.717 & 0.913 & 0.807\\
			HONE & 0.624 & 0.535 & 0.709 & 0.609 & 0.937 &\textbf{ 0.841}\\
			SCAT & 0.812 & \textbf{0.930} &  0.718 & 0.826 & 0.931 & 0.815 \\
			RGCN & 0.808 & 0.848 & 0.721 & 0.786 & 0.883 & 0.805\\
			\midrule
			Common Neighbors (CN) & 0.631 & 0.818 & 0.711 & 0.679 & 0.881 & 0.732 \\
			Jaccard's Coefficient (JC) & 0.624 & 0.816 & 0.702 & 0.678 & 0.883 & 0.733 \\
			Adamic-Adar (AA) & 0.637 & 0.819 & 0.715 & 0.68 & 0.884 & 0.725\\
			\bottomrule
		\end{tabular}
	\end{table*}
	\begin{table*}[t]
		\centering
		\makeatletter\def\@captype{table}\makeatother
		\caption{ Influences of motif types on AUC} \label{table:5}
		\begin{tabular}{c|c|c|c|c|c|c}
			\toprule
			Motif type & Youtube & LiveJournal & Bio-sc-cc & Bio-sc-ht & DBLP & Ca-GrQc \\
			\midrule
			$M_{31}$ & 0.826 & 0.911 & 0.781 & 0.840 & 0.937 & 0.813 \\
			$M_{32}$ & 0.832 & 0.895 & 0.786 & 0.819 & 0.913 & 0.793\\
			\midrule
			$M_{41}$ & 0.838 & 0.912 & 0.786 & \textbf{0.841} & 0.833 & 0.819 \\
			$M_{42}$ & 0.819 & \textbf{0.919} & 0.777 & \textbf{0.841} & 0.942 & \textbf{0.820} \\
			$M_{43}$ & \textbf{0.843} & 0.902 & \textbf{0.793} & 0.832 & 0.931 & 0.802 \\
			$M_{44}$ & 0.833 & 0.913 & 0.787 & 0.840 & \textbf{0.944} & 0.807 \\
			$M_{45}$ & 0.839 & 0.899 & 0.787 & 0.824 & 0.909 & 0.781 \\
			$M_{46}$ & 0.824 & 0.891 & 0.755 & 0.821 & 0.899 & 0.787\\
			\bottomrule
		\end{tabular}
	\end{table*}
	\begin{table*}[t]
		\centering
		\makeatletter\def\@captype{table}\makeatother
		\caption{Average number of motifs each vertex resides in} \label{table:6}
		\begin{tabular}{c|c|c|c|c|c|c}
			\toprule
			Motif type & Youtube & LiveJournal & Bio-sc-cc & Bio-sc-ht & DBLP & Ca-GrQc \\
			\midrule
			$M_{31}$ & 198.23 & 357.78 & 323.545 & 403.01 & 29.58 & 36.56 \\
			$M_{32}$ & 11.83 & 45.78 & 50.12 & 82.33 & 13.15 & 22.14\\
			\midrule
			$M_{41}$ & 6733.23 & 18058.09 & 15118.55 & 28366.22 & 519.67 & 1073.97\\
			$M_{42}$ & 55,003.23 & 25,824.65 & 43,435.38 & 50,762.38 & 946.32 & 1,550.54 \\
			$M_{43}$ & 101.78 & 665.07 & 359.82& 738.77 & 34.97 & 66.08 \\
			$M_{44}$ & 4,768.39 & 9,777.80 & 13,718.34 & 29,609.21 & 1,063.26 & 3,002.26 \\
			$M_{45}$ & 39.11 & 155.38 & 179.61 & 308.36 & 43.99 & 82.43 \\
			$M_{46}$ & 14.81 & 16.64 & 163.02 & 204.42 & 47.13 & 174.10\\
			\bottomrule
		\end{tabular}
	\end{table*}

	\subsection{Experimental Results}
	In this section, we use AUC and precisionK to compare our proposed algorithm with nine baseline algorithms over six networks. The experimental results for AUC and precisionK are shown in Table \ref{table:4} and Fig. \ref{figure:5},  respectively. In Table \ref{table:4}, the best performance is highlighted in bold. \par
	
	\begin{figure*}[t]
		\centering
		\subfloat[Youtube]{
			\includegraphics[scale=0.25]{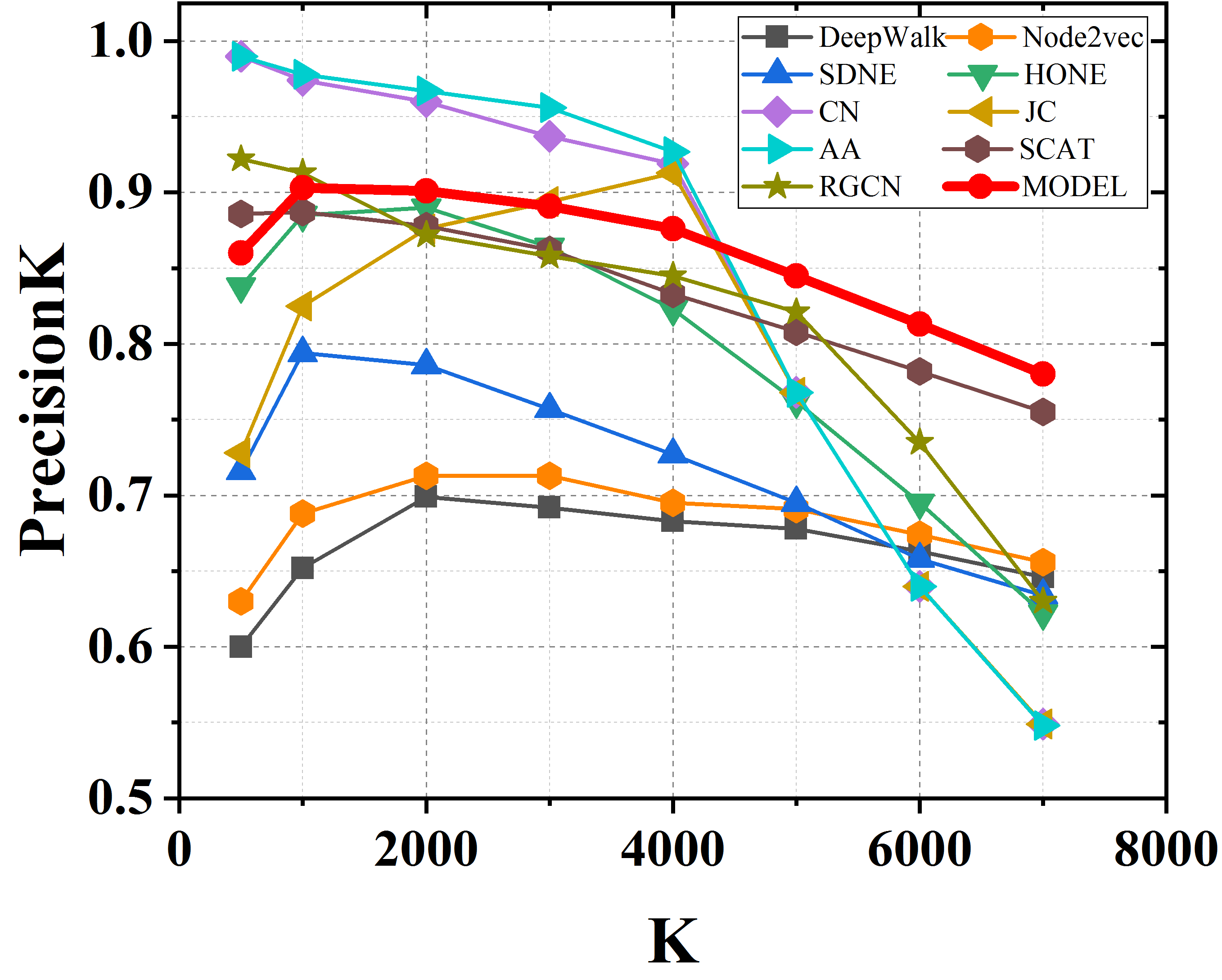}
			\label{fig5a}
		}
		\subfloat[LiveJournal]{
			\includegraphics[scale=0.25]{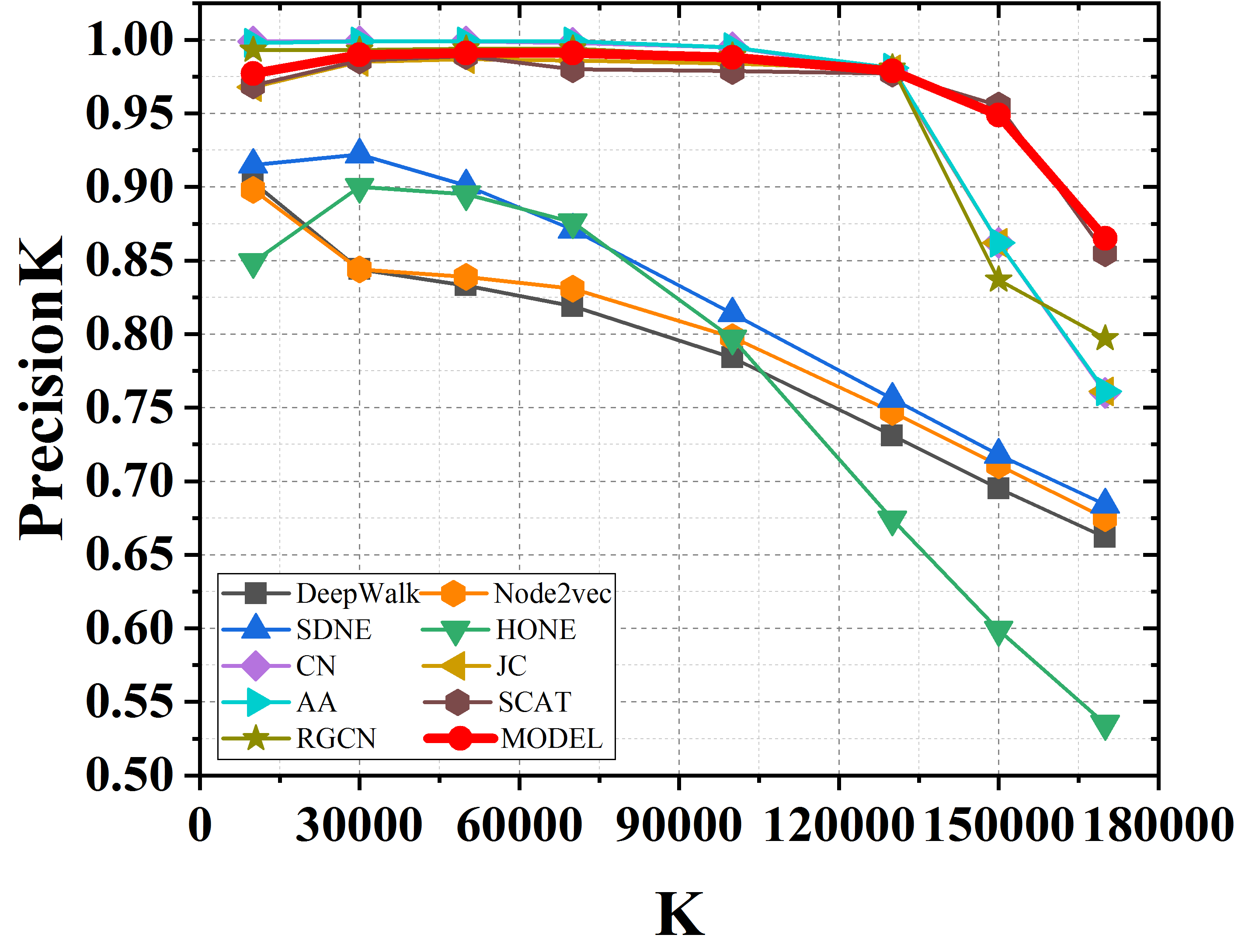}
			\label{fig5b}
		}
		\subfloat[Bio-sc-cc]{
			\includegraphics[scale=0.25]{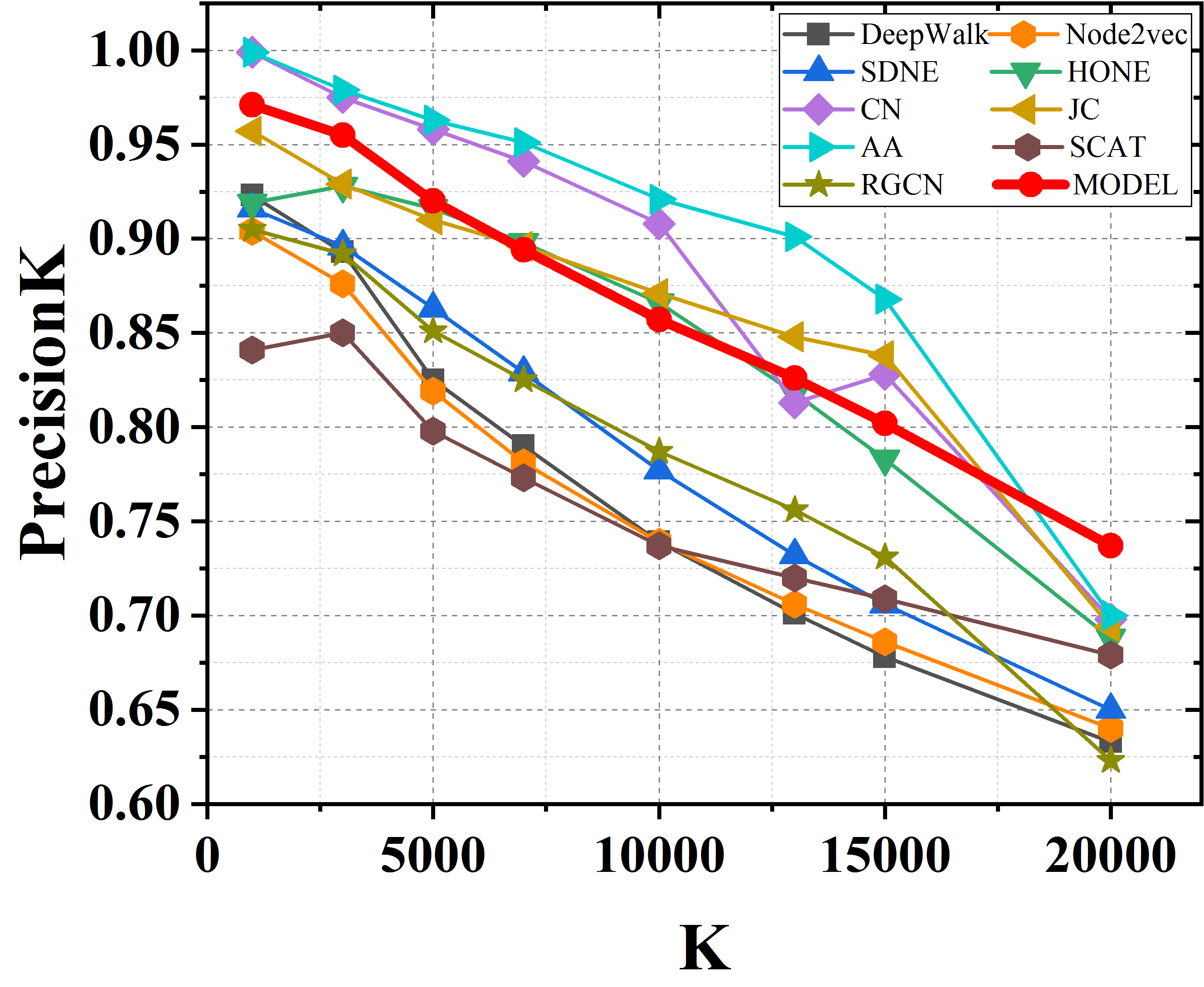}
		}
		\hfil
		\subfloat[Bio-sc-ht]{
			\includegraphics[scale=0.25]{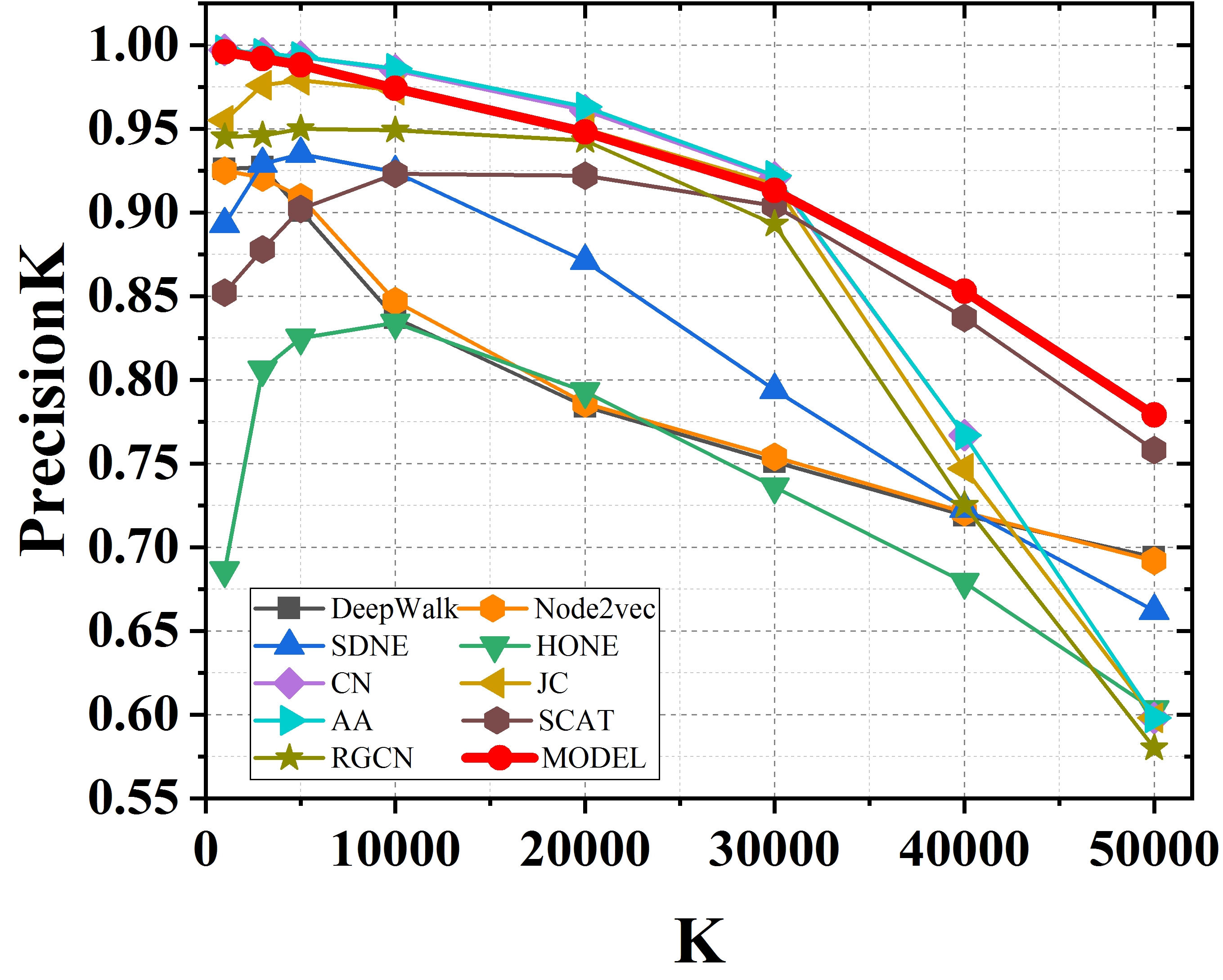}
		}
		\subfloat[DBLP]{
			\includegraphics[scale=0.25]{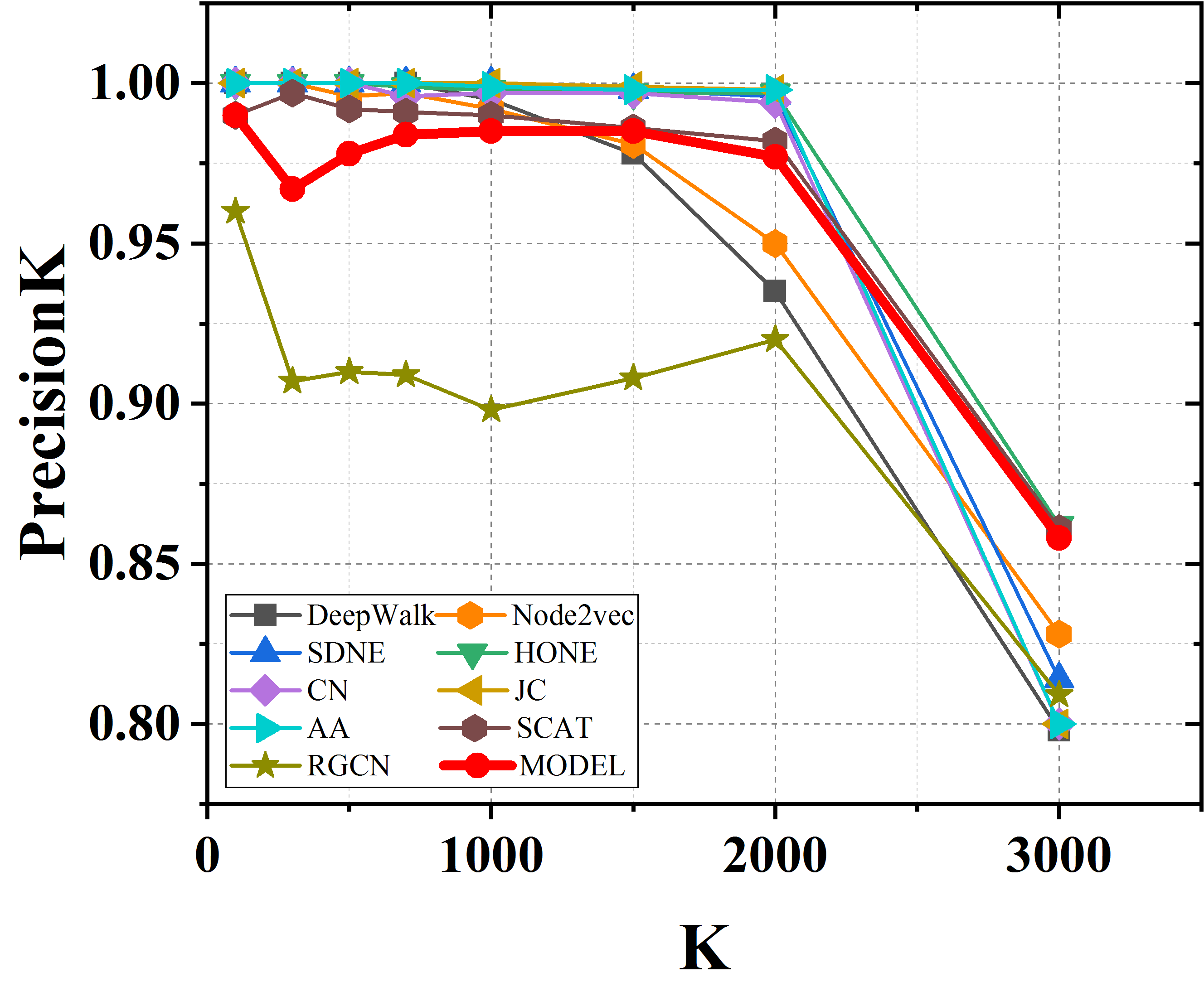}
		}
		\subfloat[Ca-GrQc]{
			\includegraphics[scale=0.25]{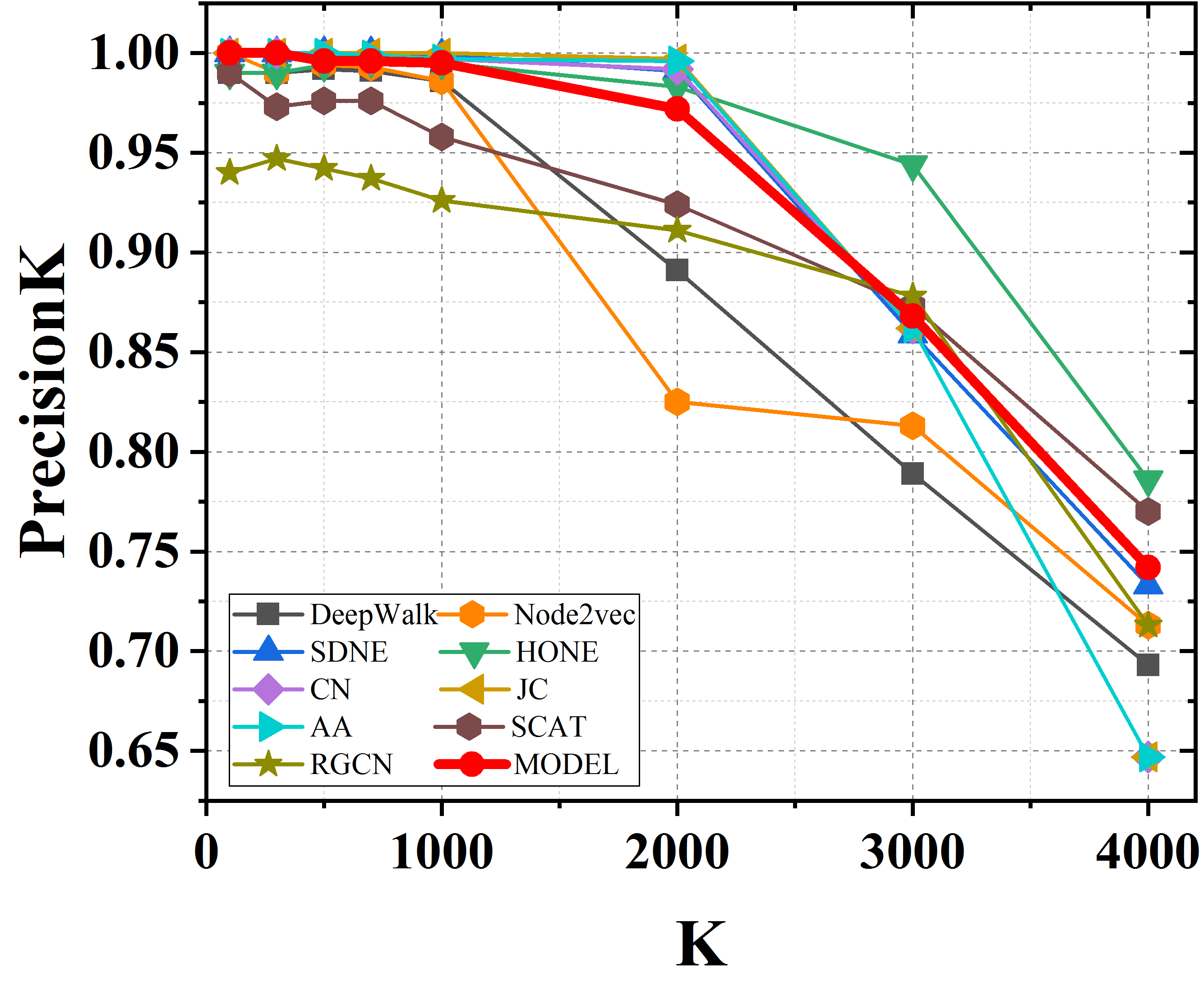}
		}
		\caption{Comparisons of ten algorithms on PrecisionK.}
		\label{figure:5}
	\end{figure*}
	\par
	\begin{figure*}[t]
		\centering
		\subfloat[Youtube]{
			\includegraphics[scale=0.25]{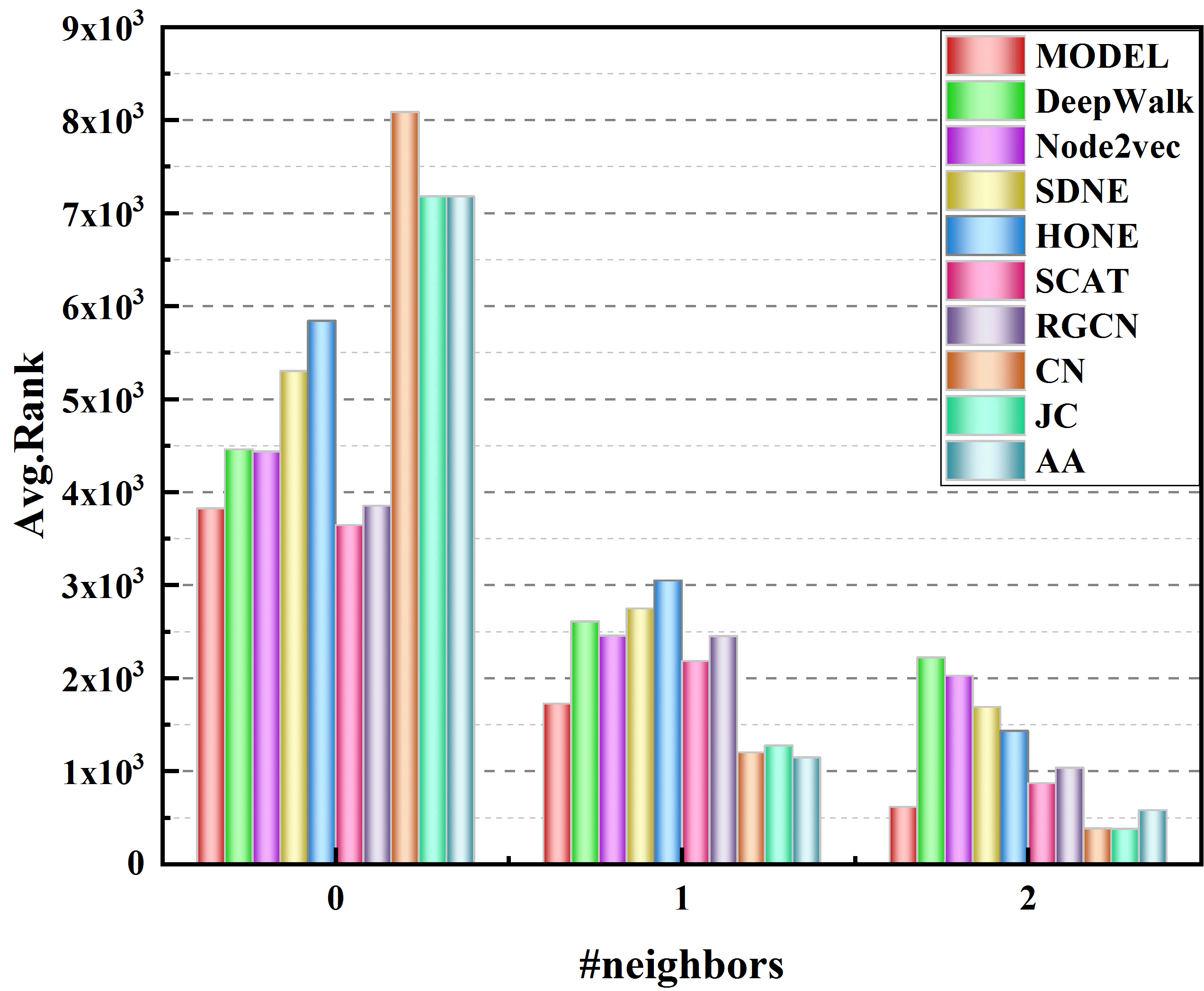}
		}
		\subfloat[LiveJournal]{
			\includegraphics[scale=0.25]{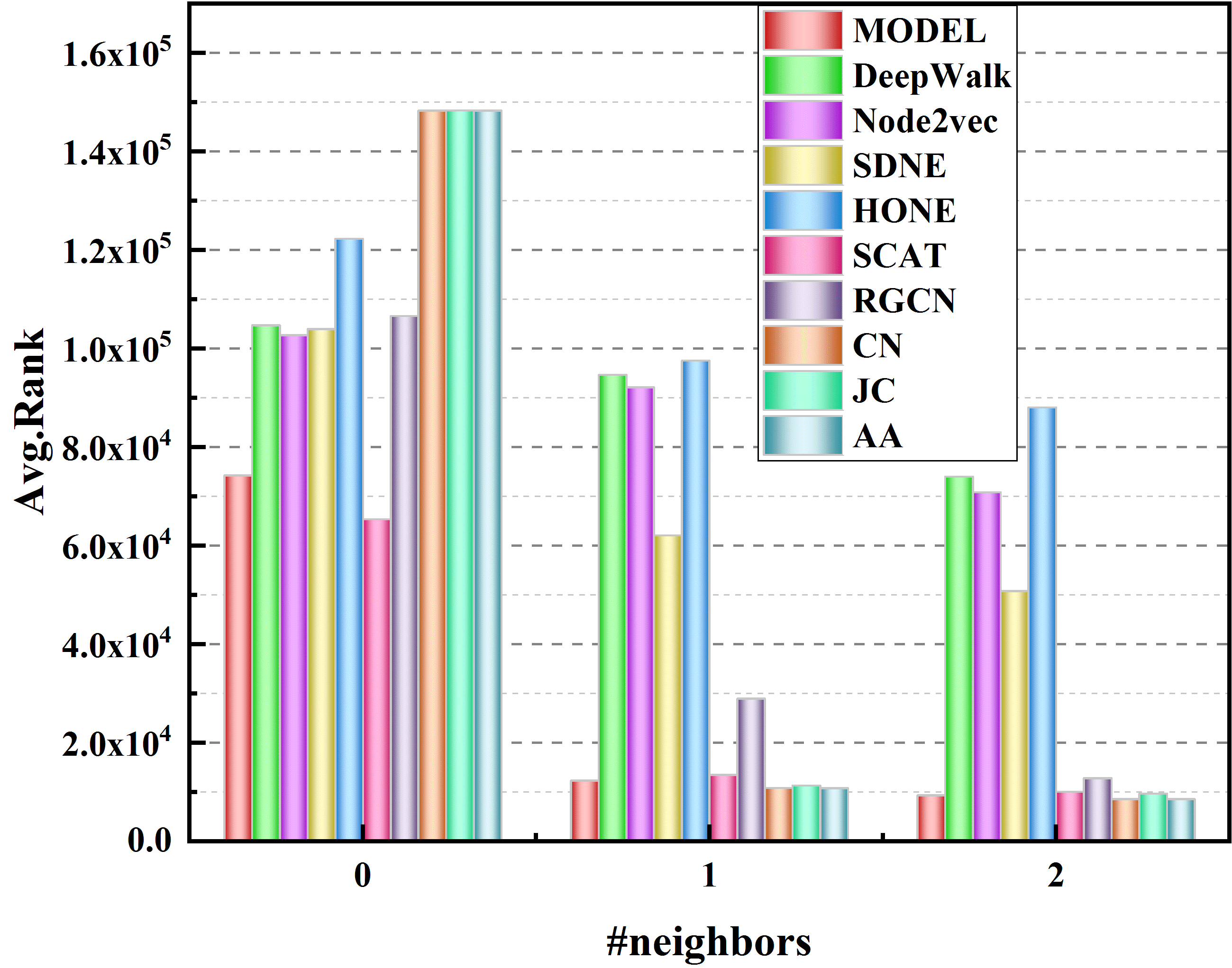}
		}
		\subfloat[Bio-sc-cc]{
			\includegraphics[scale=0.25]{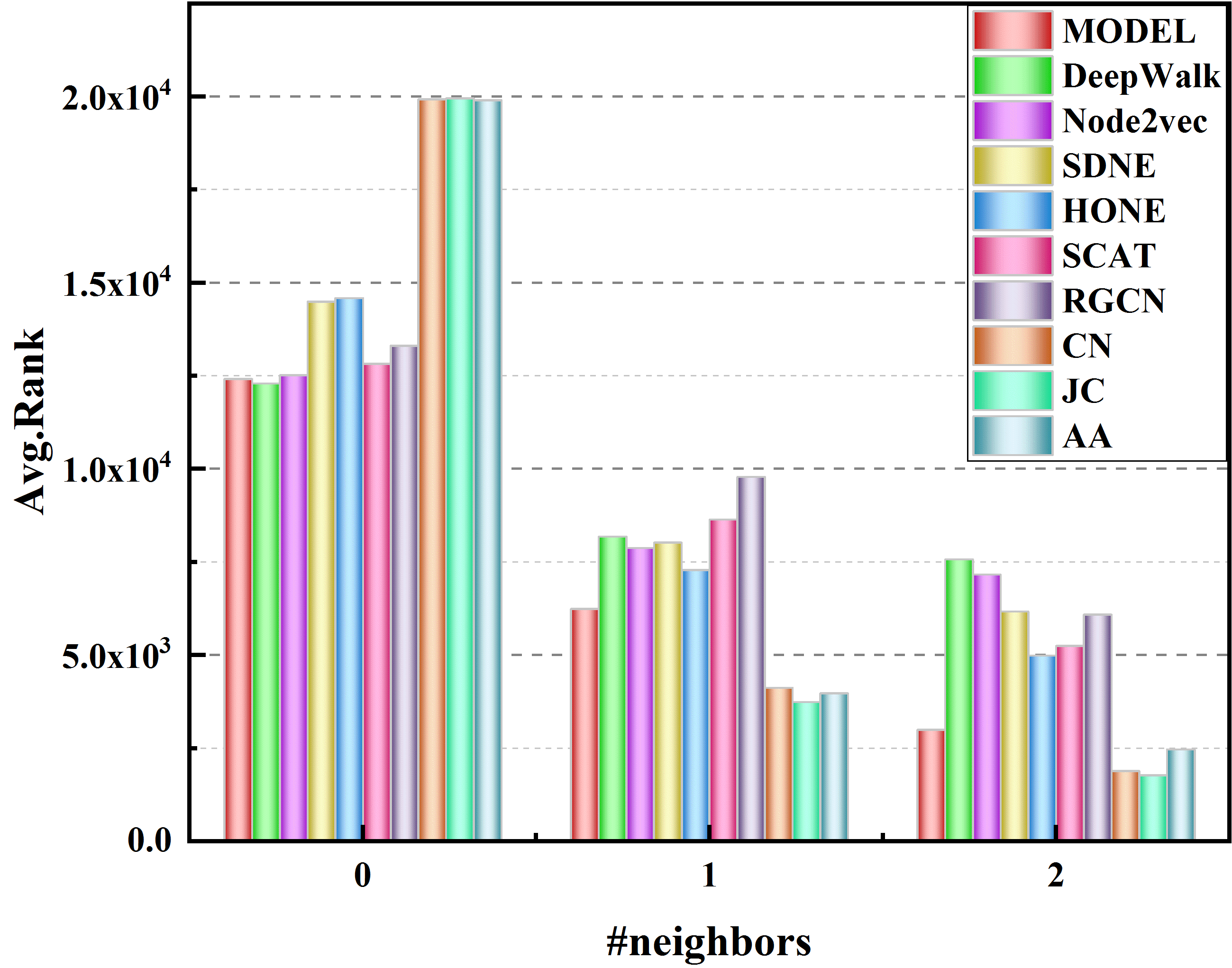}
		}
		\hfil
		\subfloat[Bio-sc-ht]{
			\includegraphics[scale=0.25]{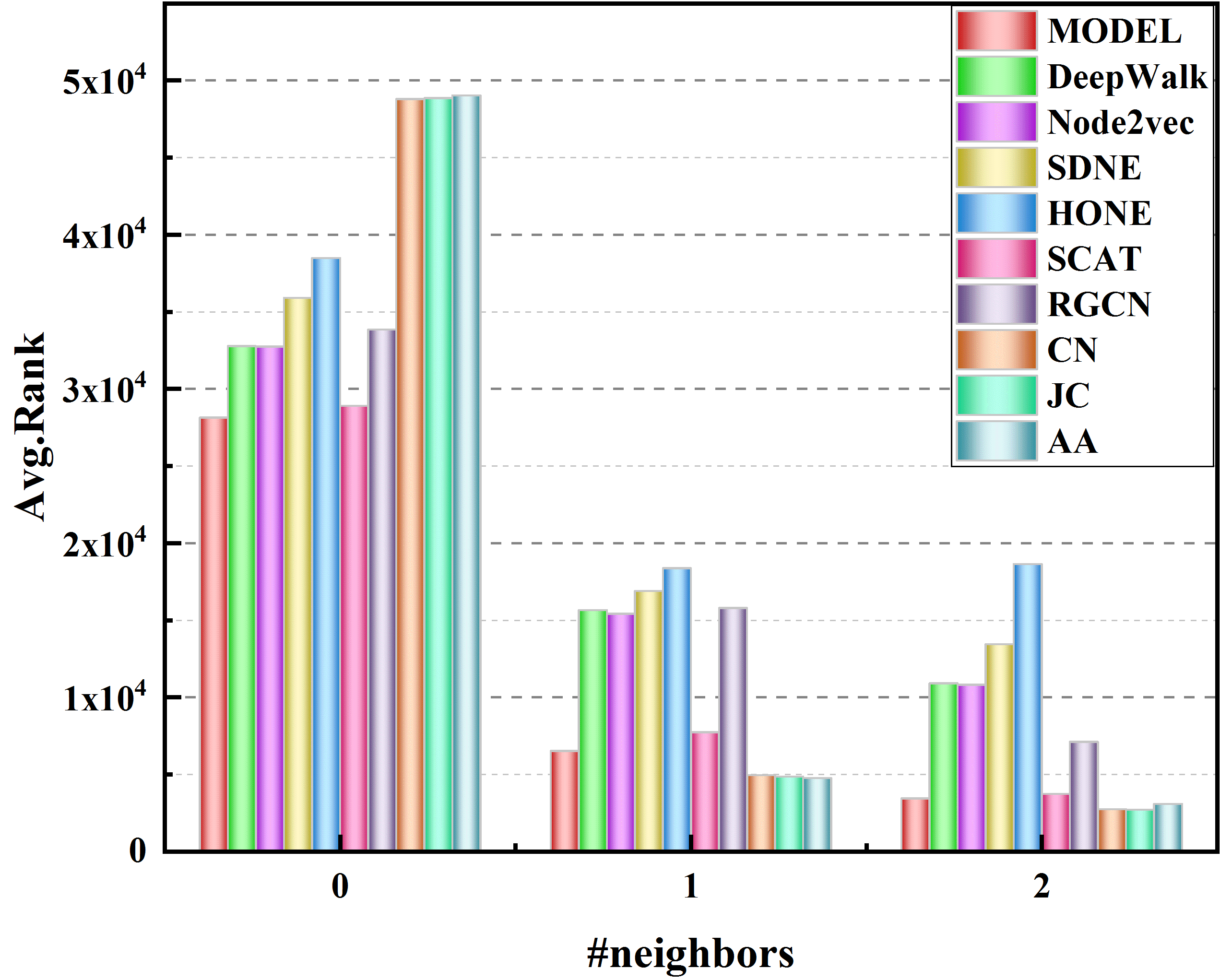}
		}
		\subfloat[DBLP]{
			\includegraphics[scale=0.25]{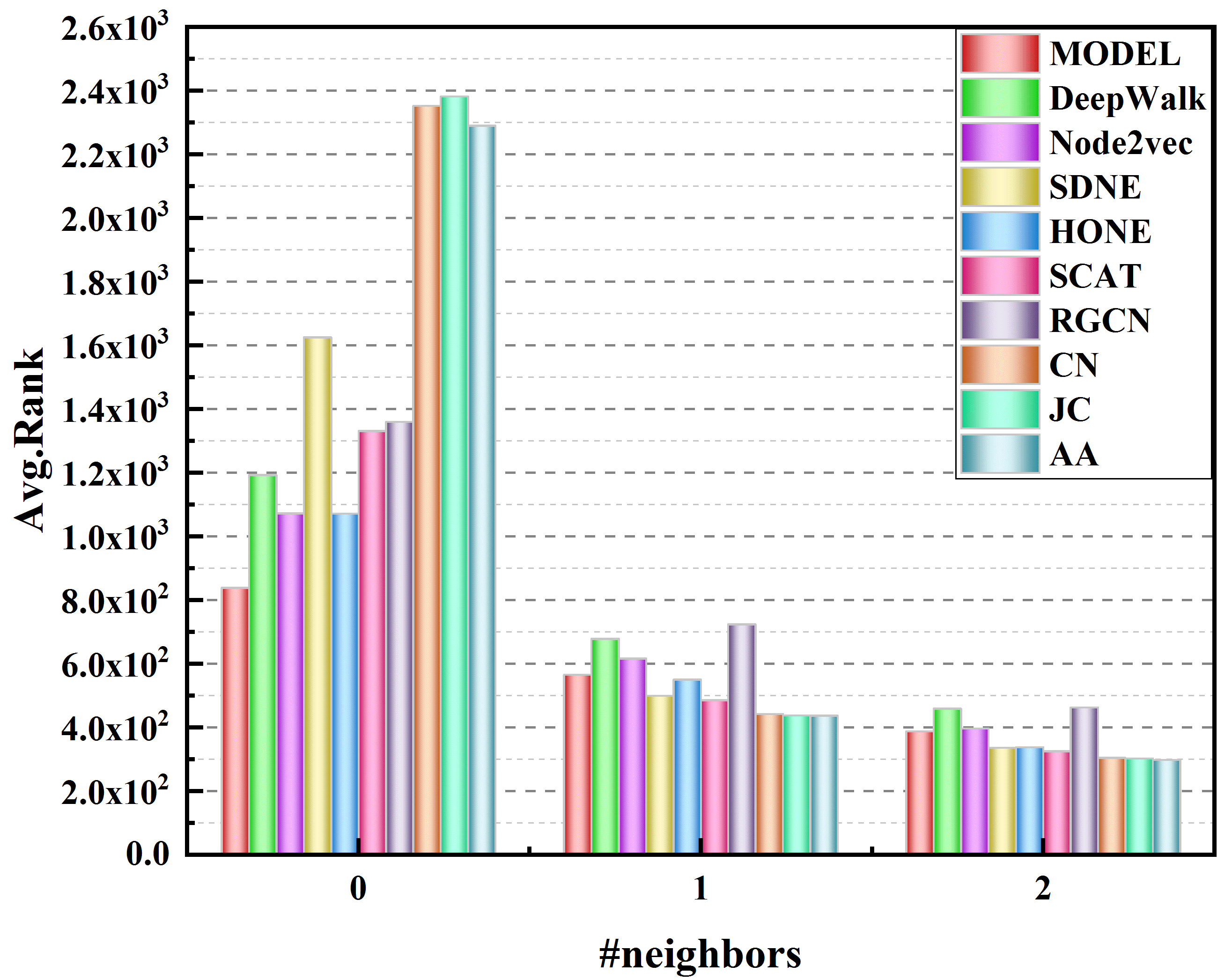}
		}
		\subfloat[Ca-GrQc]{
			\includegraphics[scale=0.25]{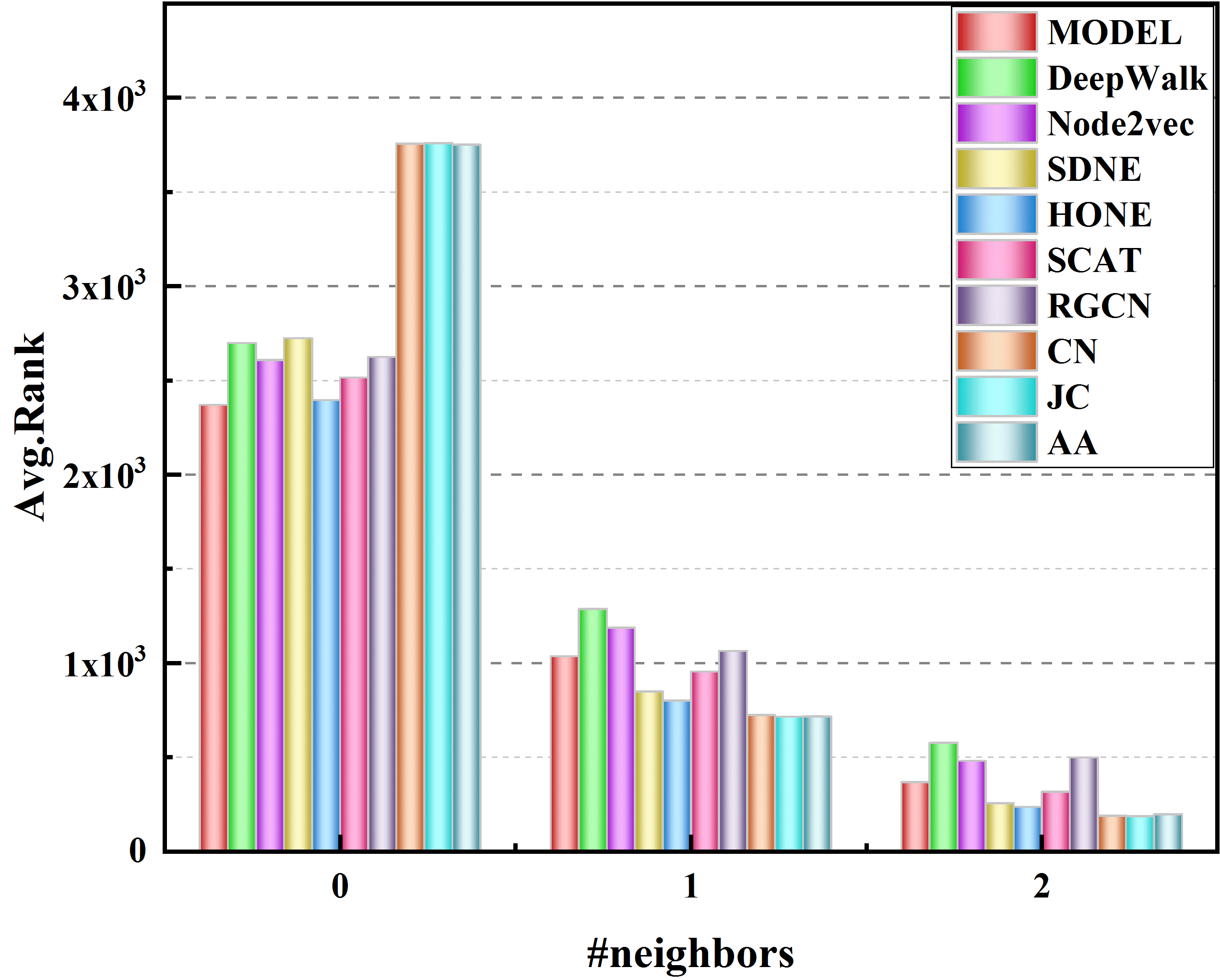}
		}
		\caption{Avg. Rank of potential weak ties.}
		\label{figure:7}
	\end{figure*}\par
	Table \ref{table:4} shows that our algorithm, MODEL, outperforms other related algorithms in most cases. Furthermore, the superiority of MODEL for percisionK is shown in Fig.  \ref{figure:5}. On the two social networks Youtube and LiveJournal, MODEL outperforms these embedding-based algorithms which do not consider motifs by up to 18.1\% and 19.5\% for AUC, respectively. Moreover, as K increases, MODEL consistently exceeds these algorithms to a large percentage for percisionK. On the two social networks, though considering motifs, HONE has the worst performance in the eight algorithms, particularly on LiveJournal. From Fig.  \ref{figure:5}, we can see that when K is relatively small, HONE has a good performance, particularly on Youtube. With K increasing, the precisionK of HONE drastically drops. On the two biological networks Bio-sc-cc and Bio-sc-ht, MODEL obtains a maximal gain of 11.6\% and 12.4\% over these embedding-based algorithms which do not consider motifs, respectively. Besides, MODEL still consistently outperforms the three algorithms for precisionK, which is evidenced by (b) and (c) in Fig. \ref{figure:5}. We also notice that HONE outperforms DeepWalk, Node2vec and SDNE, but is inferior to MODEL. In the two academic networks DBLP and Ca-GrQc, the maximal gain over embedding-based algorithms is 6.1\% and 5.9\% respectively, which is lower than that of social networks and biological networks. On the two networks, MODEL is not that superior to HONE. \par
	Table  \ref{table:4} and Fig.  \ref{figure:5} also show that embedding-based algorithms are comparable to traditional similarity-algorithms for link prediction. Traditional algorithms achieve better performance on LiveJournal and Bio-sc-cc, compared with embedding-based algorithms. It is worth noting that MODEL outperforms the three traditional algorithms over the six networks.
	
	One noticeable finding is that traditional algorithms outperform embedding-based algorithms when K is not very big. Traditional algorithms measure the similarity between two vertices by their neighbors. According to homophily \cite{mcpherson2001birds}, the similarity from neighbors greatly reflects whether two vertices are potentially connected or not. As a result, traditional algorithms give two vertices that share some common neighbors a high similarity score, whereby they can predict accurately some links. However, when vertices share few or no common neighbors, traditional algorithms cannot work. For example, common neighbors give two vertices that are connected but have no common neighbors a similarity score of zero. Thus, traditional algorithms only can predict a portion of links, which explains the phenomenon that their performance abruptly becomes poor when K exceeds a threshold. Examples include Youtube, where their PrecisionKs drop drastically when K is over 4000. Embedding-based algorithms use overall network structure, keeping better balance between vertices that have some and others that have no common neighbors. Although embedding-based algorithms may not give a higher score to two vertices that have more common neighbors, they give a reasonable score to two vertices that have no common neighbors. Thus, Embedding-based algorithms outperform traditional algorithms in most cases when K is the maximum value. 
	\subsection{Influence of Motif Types}
	In the above section, we show the superiority of our proposed algorithm. In this section, we investigate the influence of motif types on the six networks. The experimental results for AUC are shown in Table  \ref{table:5} and the best performance is highlighted in bold.\par
	Table \ref{table:5} shows that the most appropriate motif type is one among $M_{42}$, $M_{43}$, and $M_{44}$. For the six networks, $M_{32}$, $M_{45}$, and $M_{46}$ always achieve bad performance. The reasons may be as follows: 1) From Fig. \ref{figure:2}, we notice that $M_{32}$, $M_{45}$, and $M_{46}$ are close-knit cliques, and even there is not another edge in $M_{32}$ and $M_{46}$. As a result, leveraging them cannot better capture  missing edges than leveraging sparser cliques. 2) The average number of motifs that each vertex resides in can be related to experimental results. To validate the assumption 2, we conduct an experiment to investigate the average number of motifs each vertex resides in. The result is shown in Table \ref{table:6}. From Table  \ref{table:6}, we can see that the three smallest number is $M_{32}$, $M_{45}$, and $M_{46}$ in social networks and biological networks. In academic networks, the number of these three motif types is more than $M_{43}$, but less than $M_{42}$ and $M_{44}$.
	From the above two findings, we can conclude that the influence of motifs on experiment is related to their sparsity and number.
	
	From the above experiments, we know that the number and sparsity of a motif type have an impact on performance. Therefore,  we can use a small number of vertices to estimate the number of a sparse motif, such as $M_{42}$, thereby finding the most appropriate motif type.
	\begin{figure}[htbp]
		\centering
		\subfloat[\#dimension]{
			\includegraphics[scale=0.3]{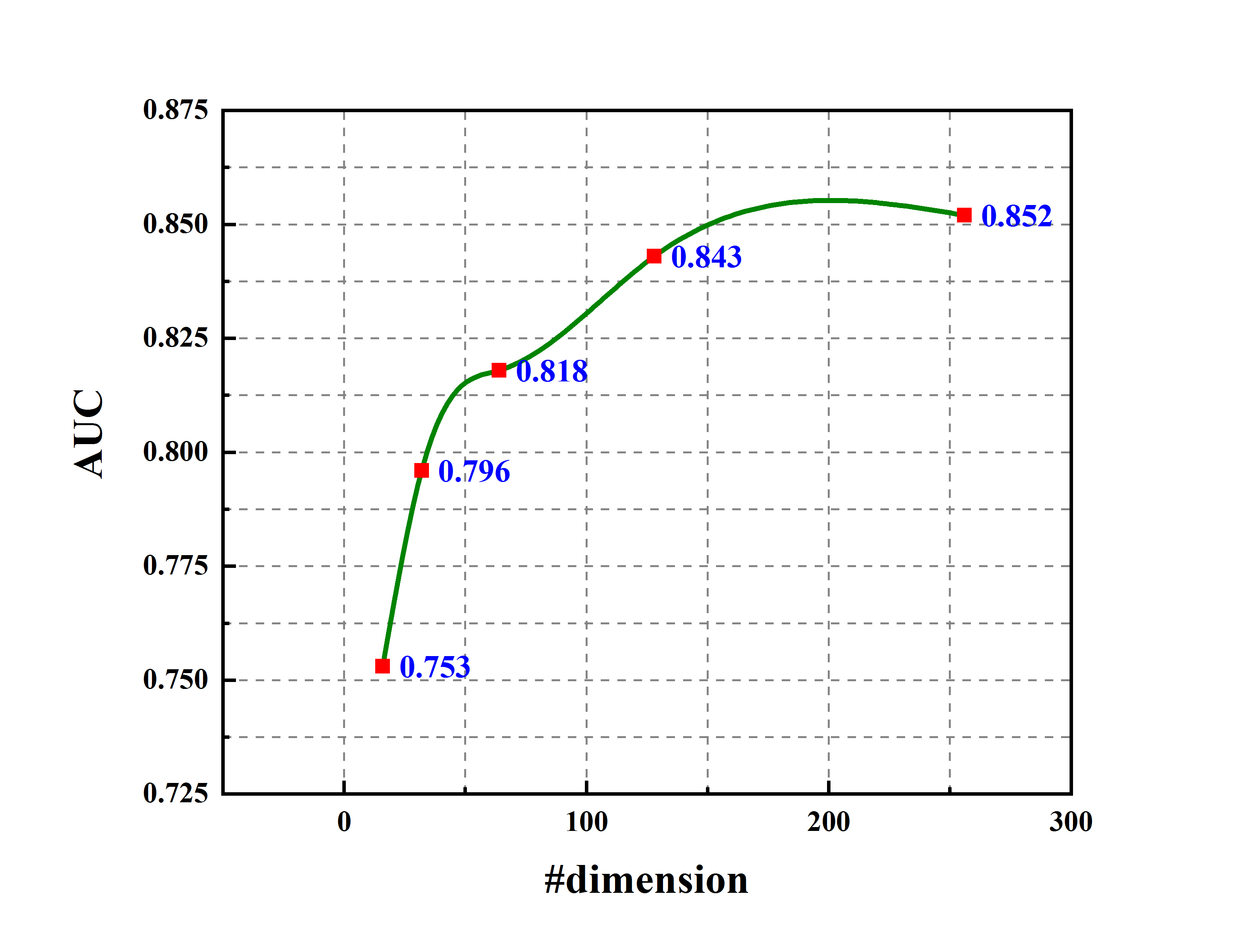}
			\label{fig6a}
		}
		\hfil
		\subfloat[$\alpha$]{
			\includegraphics[scale=0.3]{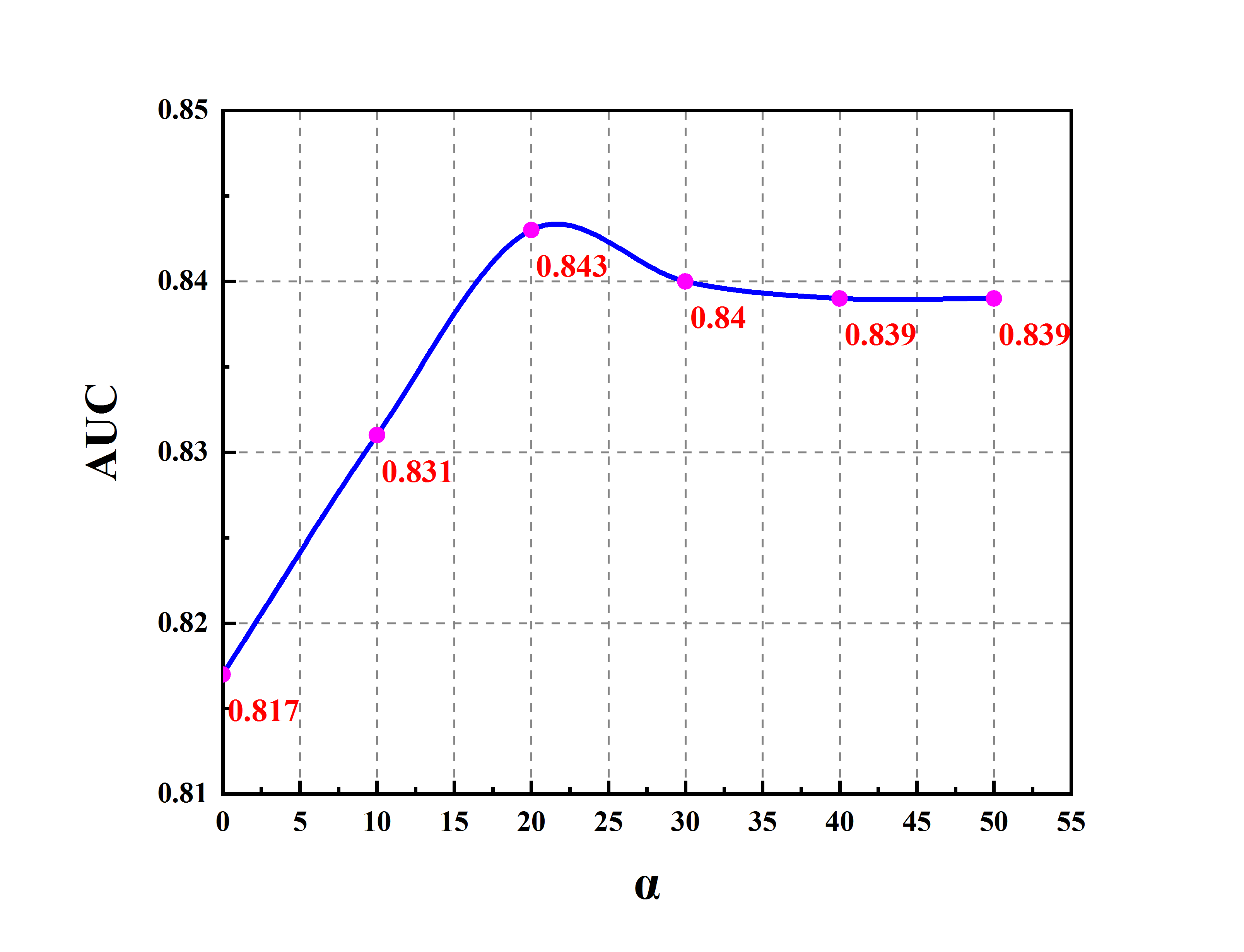}
			\label{fig6b}
		}
		\caption{Sensitivity w.r.t. dimension and $\alpha$}
		\label{figure:6}
	\end{figure}\par
	\subsection{Weak Ties}
	For link prediction, a serious problem is how to effectively predict potential weak ties as their two endpoints share a small number of common neighbors. However, the problem is always ignored by existing algorithms.
	Defining the strength of an edge is beyond the scope of this study, but we can use the number of common neighbors to define the strength of an edge according to \cite{onnela2007structure}.\par
	
	To evaluate the effectiveness in predicting potential weak ties, we first select all hidden edges whose two endpoints share a few common neighbors as positive examples.
	In this paper, a hidden egde is selected if the number of neighbors shared by its two endpoints is less than three. The new test examples consist of the selected positive examples and all negative examples used for experiments in Section \textit{Experimental Results}. We sort the scores of elements in the new test examples by decreasing, and report the average rank of selected positive examples. A smaller average rank implies a better prediction of weak ties. We show the results in Fig. \ref {figure:7}.\par
	
	From Fig. \ref{figure:7}, we can see that when the number of common neighbors is zero, embedding-based algorithms  achieve the best performance, and three traditional similarity-based algorithms, Common Neighbors (CN), Jaccard's Coefficient (JC) and Adamic-Adar (AA), achieve always the worst performance. Moveover, MODEL always ranks at the top and achieve the best performance in Bio-sc-ht, DBLP and Ca-GrQc. However, when the number is one or two, traditional algorithms all achieve the best performance.
	Even so, MODEL still achieves great performance, no big gap compared to traditional algorithms.
	The above phenomenon may be due to that traditional algorithms only focus on direct neighbors to derive the similarity of two vertices. Thus, when the number of common neighbors is zero, the derived similarity of potentially connected vertices is zero, which is the same as the similarity of most unconnected vertices. MODEL addresses the problem of no common neighbors by redefining vertices' neighbors, thus it significantly outperforms traditional algorithms when no common neighbors exist. If two vertices have common neighbors, the number of common neighbors can be a good indicator for predicting edges. Fig. \ref{figure:7} also shows that MODEL outperforms embedding-based algorithms regardless of the number of common neighbors. \par

	\subsection{Parameter Sensitivity}
	In this section, we investigate the parameter sensitivities of $\alpha$ and the number of dimensions, and report how the two hyper-parameters affect experimental results. we select a network Youtube and motif type $M_{43}$ to conduct experiments. The experimental results are shown in Fig. \ref {figure:6}.

	\textbf{The number of embedding dimensions}: In experiments, it is necessary to choose an appropriate number of embedding dimensions.
	A vector space with higher dimensions is more capable of encoding more useful information in original networks. However, it is more difficult to train models to learn higher-dimensional vector representations since time complexity is higher. In this experiment, we choose five different number of dimensions (16, 32, 64, 128, 256). From Fig. \ref{fig6a}, we can see that the value of AUC increases rapidly as the dimension goes from 16 to 128. However, when the dimension increase from 128 to 256, the value increases gently, and is insensitive to the change of dimension. Overall, we can achieve a good performance in a low-dimensional space.\par
	\textbf{Balance between the first-order proximity and the second-order proximity}: In the proposed algorithm, $\alpha$ is used to control the balance between the first-order proximity and the second-order proximity. From Fig. \ref{fig6b}, we can see that the performance of $\alpha>0$ is better than that of $\alpha=0$. This is due to the reason that we only preserve the second-order proximity when $\alpha=0$. With $\alpha$ increasing, the weight of the first-order proximity becomes large gradually, and the performance becomes better. We also see that the performance declines slightly and keeps stable at last when $\alpha>20$, which demonstrates that we must make a trade-off between the two proximities.
	\subsection{Time Complexity}
		\begin{figure}[htbp]
		\centering
		\subfloat[Bio-sc-cc]{
			\includegraphics[scale=0.3]{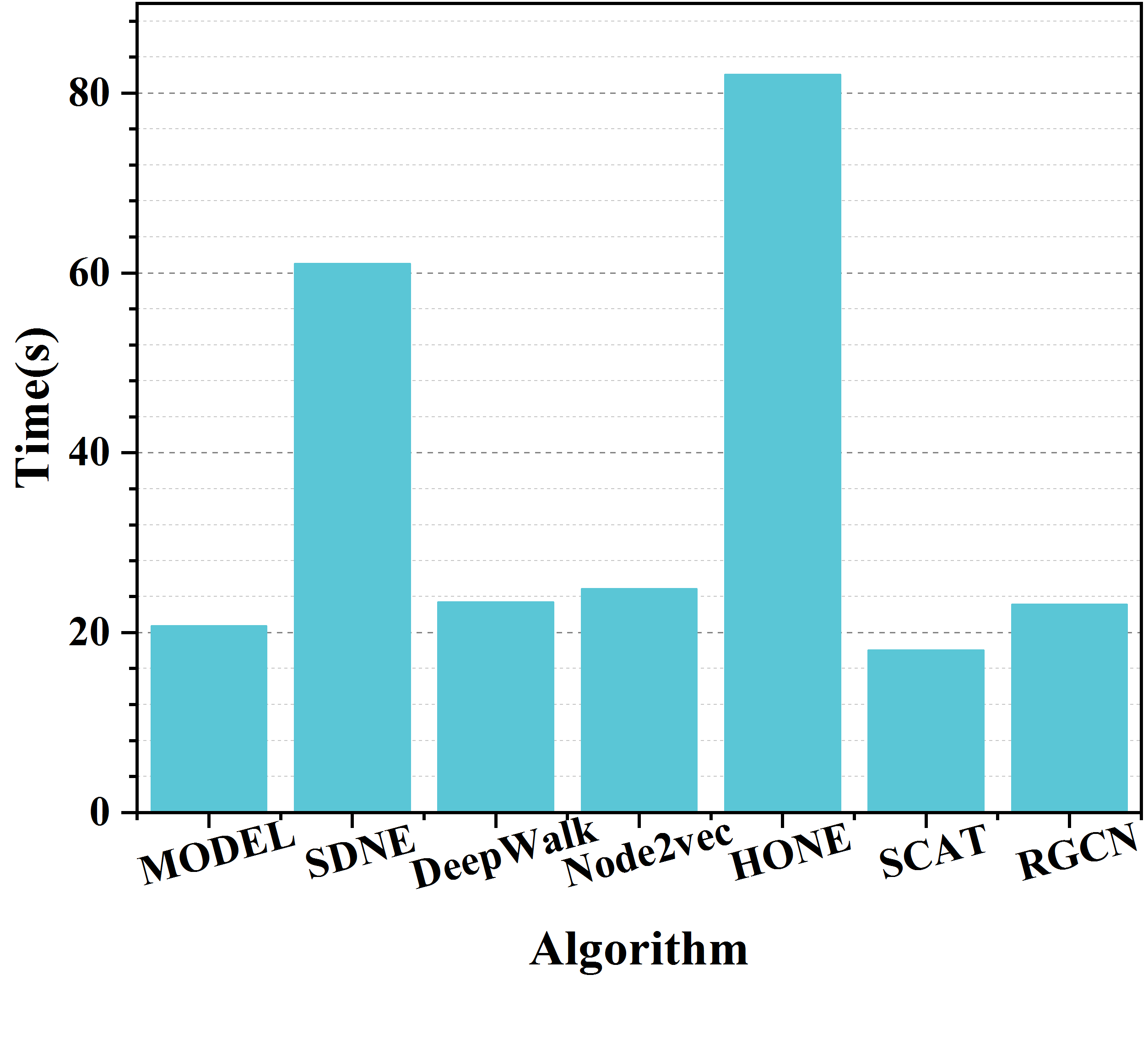}
			\label{time_a}
		}
		\hfil
		\subfloat[Youtube]{
			\includegraphics[scale=0.3]{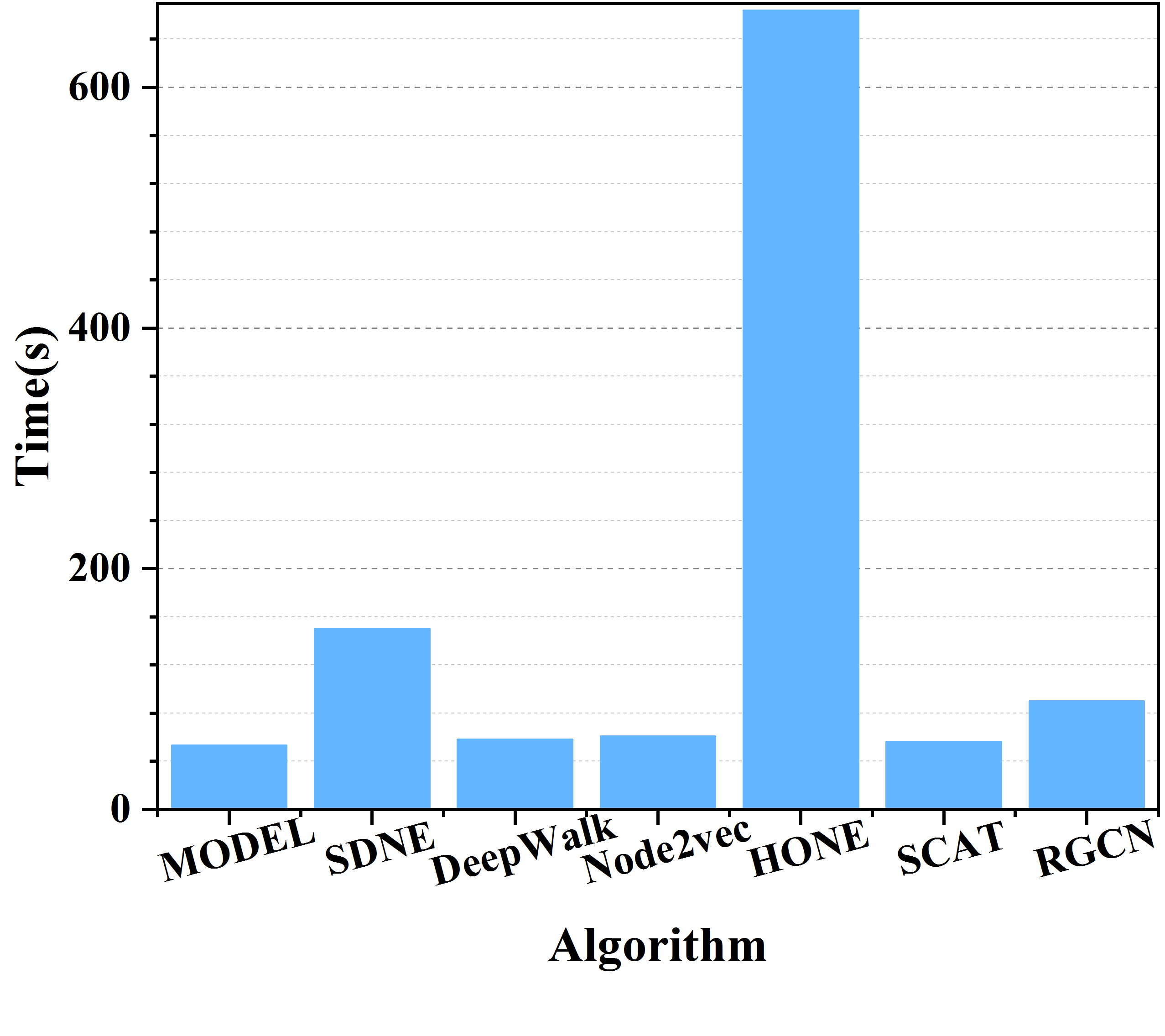}
			\label{time_b}
		}
		\caption{The running time of seven algorithms}
		\label{figure:time}
	\end{figure}\par
	 To compare the time complexity of baselines, we run experiments of the running time on a large network Youtube and a small network Bio-sc-cc. Note that we compare the running time over baselines that need to train corresponding models. Traditional algorithms are omitted here because they directly use the network structure and need no time to train a model. The results are shown in Fig. \ref{figure:time}.
	 
	 From Fig. \ref{figure:time}, it can be seen that MODEL has less running time, compared with these baselines. In Bio-sc-cc, the running time of MODEL is only more than that of SCAT, and the time of MODEL is lowest in larger network Youtube. These results suggest that MODEL achieve satisfactory performance in time complexity. We can notice that the running time of HONE is highest in the two networks. Furthermore, in Youtube it significantly exceeds the running time of SDNE, the second most time. This reason is that HONE has a great number of matrix computations and uses all two-node and four-node motifs.
}

\section{Conclusion}\label{sec6}
We proposed a motif-based algorithm to learn node embedding for link prediction.
The algorithm uses motifs, through defining two levels of proximity in autoencoders,
to capture the higher-order structure and alleviate the problem of predicting weak ties.
Experiments on six real-world networks, covering social, biological and academic networks,
demonstrate that the proposed algorithm significantly outperforms the state-of-the-art baseline algorithms.
The maximal gain on AUC over embedding-based algorithms can be up to 19\%,
and up to 20\% over traditional similarity-based algorithms.
In addition, we have also investigated the influence of motif types on the link prediction performance,
which hasn't studied before.
From the experimental results, we obtained insights on how to select the most
appropriate motif type for a specific task in networks.

In the future, we will consider sampling importance motifs so as to apply this method to large networks.



\ifCLASSOPTIONcaptionsoff
  \newpage
\fi

	\bibliographystyle{IEEEtran}
\bibliography{IEEEabrv,reference}


%

%

\begin{IEEEbiography}[{\includegraphics[width=1in,height=1.5in,clip,keepaspectratio]{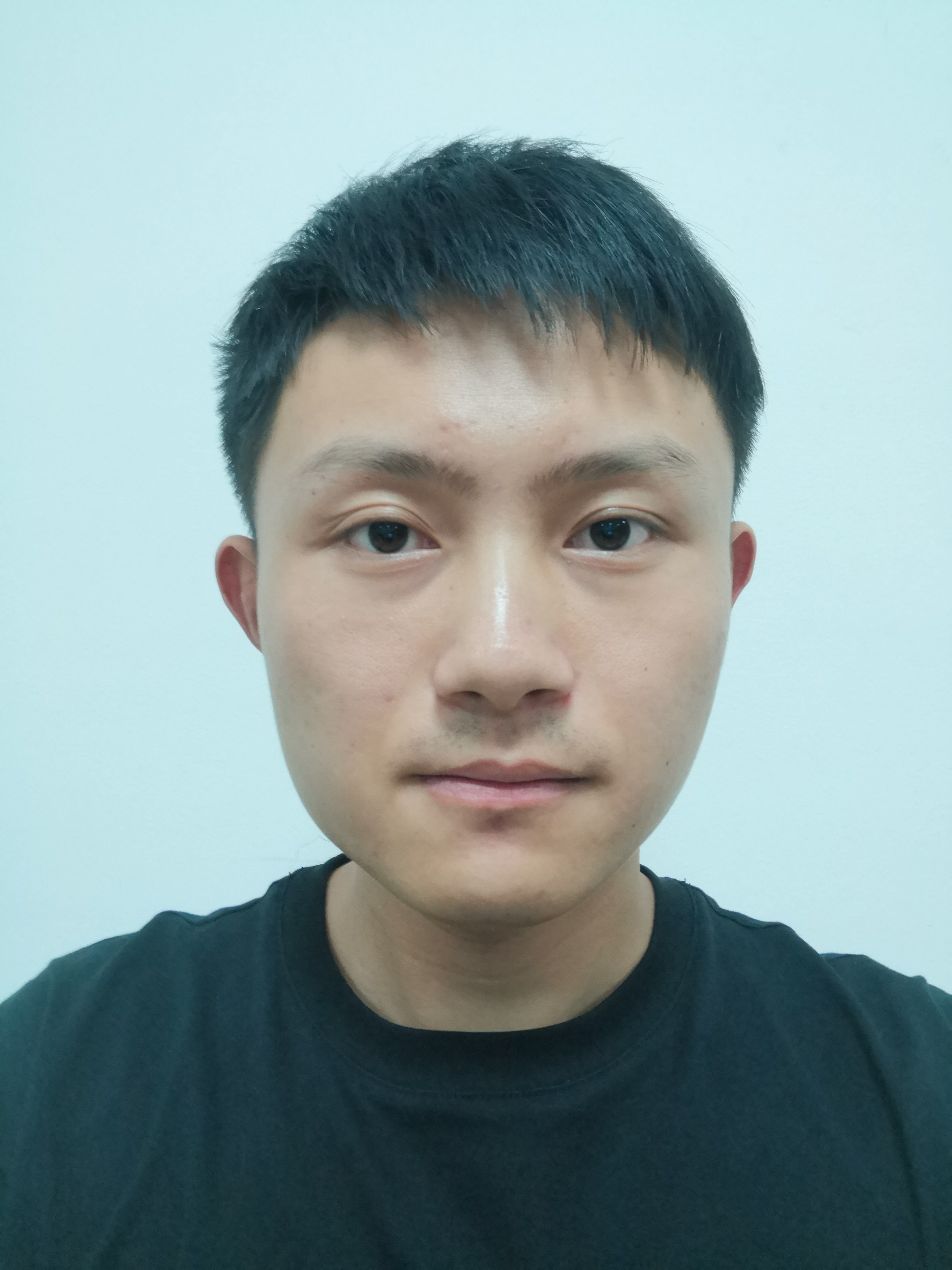}}]{Lei Wang} received the BSc degree in software engineering from Dalian University of Technology, China, in 2018. He is currently	working toward the master’s degree in the School of Software, Dalian University of Technology, China. His research interests include data mining, analysis of complex networks, and machine learning.

\end{IEEEbiography}
\begin{IEEEbiography}[{\includegraphics[width=1in,height=1.5in,clip,keepaspectratio]{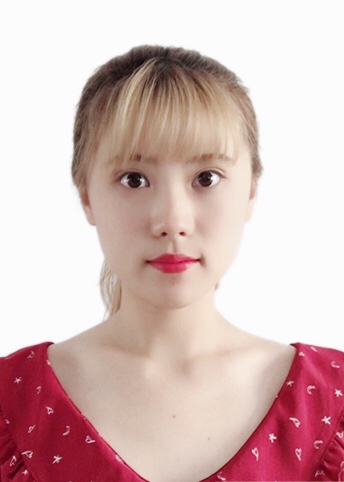}}]{Jing Ren} received the BSc degree in software engineering from Huaqiao University, Xiamen, China. She is currently pursuing the master's degree in software engineering in Dalian University of Technology, China. Her research interests include big scholarly data, network science, and computational social science.
	
\end{IEEEbiography}

\begin{IEEEbiography}[{\includegraphics[width=1in,height=1.25in,clip,keepaspectratio]{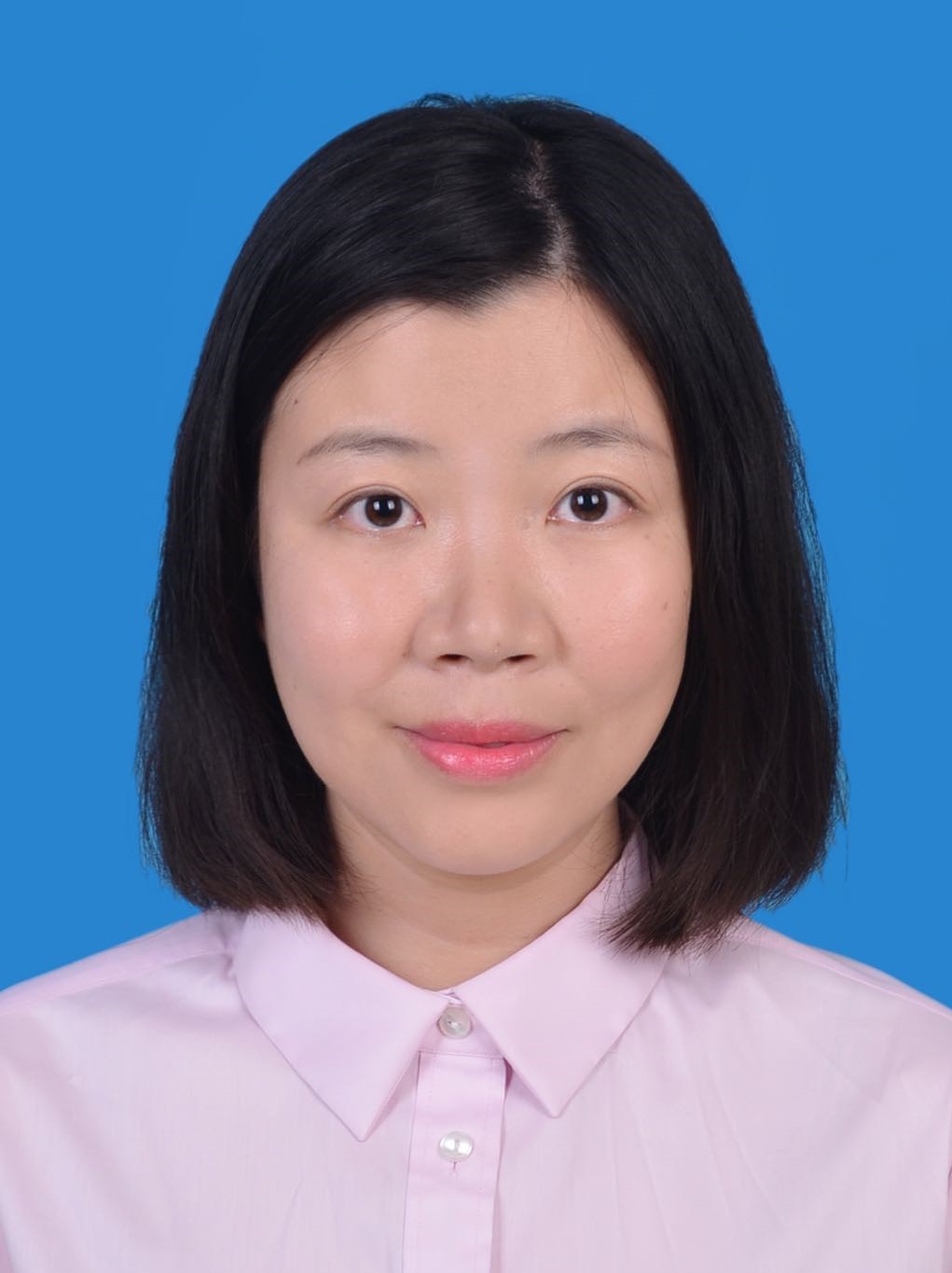}}]{Bo Xu}
received the BSc and PhD degrees from the Dalian University of Technology, China, in 2007 and 2014, respectively. She is currently an Associate Professor in School of Software at the Dalian University of Technology. Her current research interests include data science, network analysis and natural language processing.
\end{IEEEbiography}

\begin{IEEEbiography}[{\includegraphics[width=1in,height=1.25in,clip,keepaspectratio]{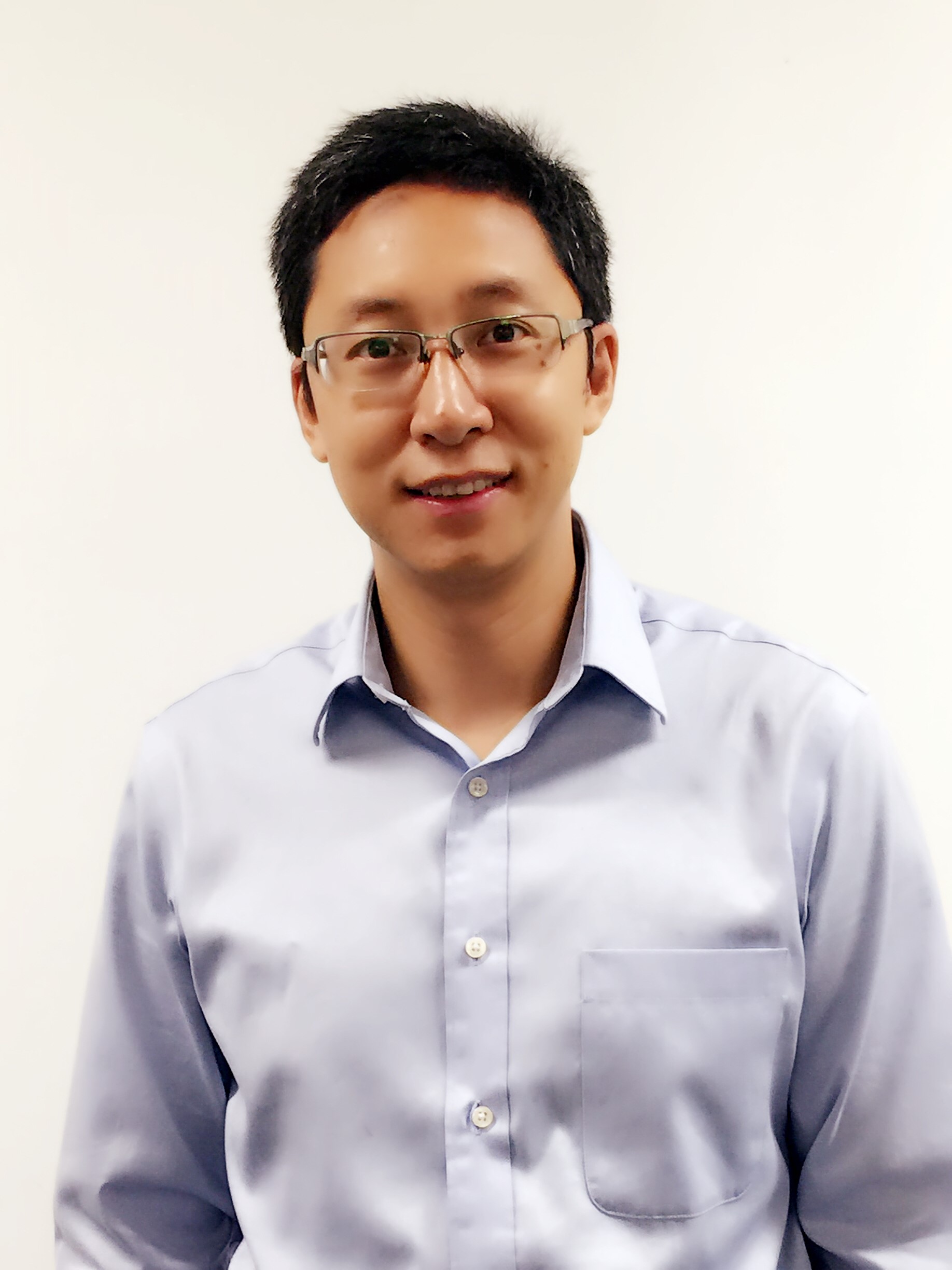}}]{Jianxin Li} received the PhD degree in computer science from the Swinburne University of	Technology, Australia, in 2009. He is an Associate	Professor in the School of Info Technology,	Deakin University. His research interests include	database query processing \& optimization, social	network analytics, and traffic network data	processing.
	
\end{IEEEbiography}
\begin{IEEEbiography}[{\includegraphics[width=1in,height=1.25in,clip,keepaspectratio]{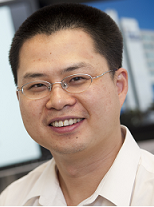}}]{Wei Luo} is a Senior Lecturer in Data Science at Deakin University. Wei's recent research focuses on machine learning and its application in health, sports, and cybersecurity. Wei has tackled a number of key information challenges in healthcare delivery and has published more than 50 papers in peer-reviewed journals and conferences. Wei holds a PhD in computer science from Simon Fraser University, where he received training in statistics, machine learning, computational logic, and modern software development.
\end{IEEEbiography}

\begin{IEEEbiography}[{\includegraphics[width=1in,height=1.25in,clip,keepaspectratio]{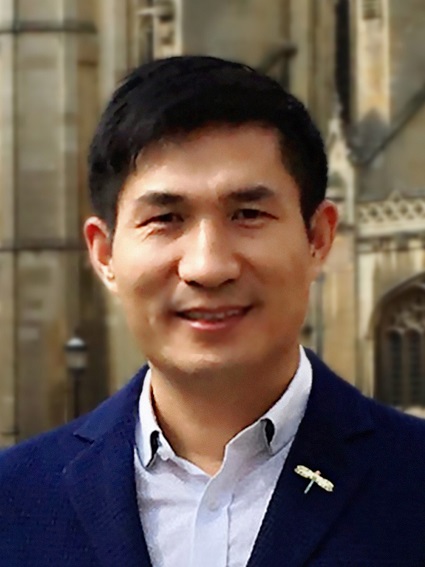}}]{Feng Xia}
Feng Xia (M’07-SM’12) received the BSc and PhD degrees from Zhejiang University, Hangzhou, China. He is currently an Associate Professor and Discipline Leader in School of Science, Engineering and Information Technology, Federation University Australia, and on leave from School of Software, Dalian University of Technology, China, where he is a Full Professor. Dr. Xia has published 2 books and over 300 scientific papers in international journals and conferences. His research interests include data science, knowledge management, social computing, and systems engineering. He is a Senior Member of IEEE and ACM.
\end{IEEEbiography}





\end{document}